

\input phyzzx.tex
\def\np{Nucl. Phys.}
\def\pl{Phys. Lett.}

\def\cmp{Comm. Math. Phys.}

\def\mpl{Mod. Phys. Lett.} \def\half{{1\over 2}}

  \def\tr{{\hbox{\rm Tr}}}

\def\mani{\cal{M}}

\def\cald{{\cal{D}}}

\def\ele{{\hbox{\sevenrm L}}}
\def\ere{{\hbox{\sevenrm
R}}} \def\lai{{\hbox{\sevenrm I}}}
\def\laf{{\hbox{\sevenrm f}}}

 \def\D{\nabla}
\def\deltat{\tilde\delta}

\def\laga{{\hbox{\sevenrm g,tYM}}}
\def\lag{{\hbox{\sevenrm g}}}
\def\la{{\hbox{\sevenrm tYM}}}
\def\ym{{\hbox{\sevenrm YM}}}

\tolerance=500000 \overfullrule=0pt

\pubnum={US-FT-8/93 \cr hep-th/9404115}
\date={March, 1994}
\pubtype={}
\titlepage

\title{TOPOLOGICAL  MATTER IN FOUR DIMENSIONS}
 \author{ M. Alvarez and J.M.F. Labastida\foot{E-mail: LABASTIDA@GAES.USC.ES} }
\address{Departamento de F\'\i sica de Part\'\i culas\break Universidade de
Santiago\break E-15706 Santiago de Compostela, Spain}

\abstract{Topological models involving matter couplings to Donaldson-Witten
theory are presented. The construction is carried  using
both, the  topological algebra and its  central extension, which  arise from
the twisting of $N=2$ supersymmetry in four dimensions.
The framework in which the
construction is based is constituted by the
superspace associated to these algebras.
The models show new
features of topological quantum field theories which could provide
either a mechanism for
topological  symmetry breaking, or the analog of two-dimensional mirror
symmetry
in four dimensions.}

\endpage \pagenumber=1


\chapter{Introduction}

Topological matter in two dimensions \REF\tsm{E. Witten\journal\cmp&118(88)411}
\REF\ey{T. Eguchi and S.K. Yang\journal\mpl&A5(90)1693} \REF\vafa{C.
Vafa\journal\mpl&A6(91)337} \REF\pablo{J.M.F. Labastida and P.M.
Llatas\journal\np&B379(92)220} [\tsm,\ey,\vafa,\pablo] have shown to be a very
interesting framework to formulate \REF\witten{E. Witten, ``Mirror Manifolds
and Topological Field Theory", in {\it Essays on Mirror Manifolds}, ed.
S.-T.Yau (International Press, 1992)} \REF\vafados{C. Vafa, ``Topological
Mirrors and Quantum Rings", in {\it Essays on Mirror Manifolds}, ed. S.-T.Yau
(International Press, 1992)} [\witten,\vafados] some problems related to
mirror manifolds \REF\mirror{L.J. Dixon, in {\it Superstrings, Unified
Theories and Cosmology 1987}, (G. Furlan et al., eds.) World Scientific, 1988,
pag. 67} \REF\mirdos{W. Lerche, C. Vafa and N.P. Warner\journal\np&B324(89)427}
\REF\mirtres{B.R. Greene and M.R. Plesser\journal\np&B338(90)15} \REF\mircua{
P. Candelas, M. Lynker and R. Schmmrigk\journal\np&B341(90)383} \REF\mircin{
P.S. Aspinwall, C.A. Lutken and G.G. Ross\journal\pl&B241(90)373}
\REF\mirseis{ P. Candelas, X.C. de la Ossa, P.S. Green and L.
Parkes\journal\np&B359(91)21 \journal\pl&B258(91)118} \REF\mirsiete{ S.
Ferrara, M. Bodner and A.C.  Cadavid\journal\pl&B247(90)25}
[\mirror-\mirsiete]. Some geometrical questions which are difficult to answer
for a given Calabi-Yau manifold can be stated as simpler geometrical questions
in terms of the corresponding mirror manifold(s). Furthermore, when an $N=2$
Landau-Ginzburg model is known for the mirror manifold(s), those questions can
be stated in terms of properties of the corresponding deformed chiral ring
whose structure is fixed by the form of the Landau-Ginzburg
potential. Topological matter models have the property that out of the
complicated structure of $N=2$ superconformal theories they extract the
information regarding the geometrical questions which arise in terms of mirror
pairs. It turns out that, after twisting $N=2$ supersymmetry \REF\tqft{E.
Witten\journal\cmp&117(88)353} [\tqft,\ey], there are two types of topological
matter models [\witten,\pablo]. These two types are described by two different
forms of topological sigma models, which are called A and B models. Given a
mirror pair ${\cal M}$ and ${\cal M}'$, the vacuum expectation values (vev) of
observables of the A topological sigma model whose target manifold  is  ${\cal
M}$ are related to the ones of the B model whose target manifold is ${\cal
M}'$ [\witten]. These vev are much harder to compute for A models than for B
models. Thus the simpler computations which can be done for B models
translate as answers to difficult geometrical questions stated in terms of
computations of vev in the A model. Furthermore, when the Landau-Ginzburg
potential associated to the Calabi-Yau manifold is known, the vev  for B models
can be stated in terms of much simpler computations in the corresponding
topological Landau-Ginzburg model [\vafa].

Vafa has recently raised the question [\vafados] of whether or not there exist
some kind of mirror phenomena in four dimensions. For example, it would be very
interesting if difficult problems as the computation of Donaldson invariants
[\tqft] could be stated in much simpler terms, as it happens in two dimensions
when a Landau-Ginzburg description is available. The motivation of the work
presented in this paper is to study this question from the point of view of
twisting $N=2$ supersymmetry. A brief account of part of the results presented
here have been reported in \REF\bts{M. Alvarez and J.M.F.
Labastida\journal\pl&B315(93)251} [\bts].

There are two equivalent approaches to understand the existence of two types of
topological models in two dimensions. One approach consists of performing two
different twists of $N=2$ supersymmetry  since in two dimensions it is possible
to twist using any of the two $U(1)$ chiral currents [\witten]. The second
approach is the result of performing one of the twists to each of the  two
$N=2$
supersymmetry matter multiplets in two dimensions [\pablo]. These multiplets
are defined from chiral and twisted chiral superfields \REF\roc{S.J. Gates, C.
Hull and M. Rocek\journal\np&B248(84)157} [\roc]. Of course, the second type
of twist, when applied to these multiplets, does not lead to new  theories.

In this paper we show that in four dimensions only one type of twist is
possible. This suggest that one should study the second approach in order to
obtain topological models, namely, one should study the twisting of different
$N=2$ multiplets which describe the same on-shell physics before the twisting.
Topological Yang-Mills in four dimensions can be thought as the result of
twisting an $N=2$ supersymmetric vector multiplet [\tqft]. Only one
representation of this vector multiplet is known and therefore it seems that in
this way no new topological model could be obtained. In this paper we report
the results of an exhaustive analysis of other possible formulations of the
vector multiplet from a superspace point of view. No new topological model
with fields of spin no higher than two has been found.

The situation is rather different when one considers $N=2$ supersymmetric
matter fields. For example, several representations are known for the $N=2$
hypermultiplet \REF\fayet{P. Fayet\journal\np&B113(76)135} \REF\sohn{M.F.
Sohnius\journal\np&B138(78)109} \REF\howe{P.S. Howe, K.S. Stelle and P.K.
Townsend\journal\np&B214(83)519} [\fayet,\sohn,\howe]. This clearly opens a
line of investigation. However, this line deviates somehow from the original
motivation of the present work, namely, the construction of a theory related
to topological Yang-Mills theory which could allow a simpler way to compute
Donaldson invariants. In order to keep ourselves within our original goal, we
consider in this work topological matter coupled to topological Yang-Mills.
This will lead to a generalization of Donaldson invariants which might well be
the ones that could possess features similar to mirror symmetry.

As reported in [\bts], topological matter coupled to topological Yang-Mills
seems to possess unexpected properties. It turns out that some of the resulting
models loose some of their topological features. This is an indication that the
mirror-like hypothesis in four dimensions, if valid, is more complicated than
in two dimensions. In fact, one of the generalizations of Donaldson invariants
seems to lead to non-topological quantities. Although it  shares many of the
properties of Donaldson invariants, it  might  have a weak dependence on the
metric of the four dimensional manifold. This fact connects with one of the
most important physical problems in topological quantum field theory, namely,
the problem of
 finding mechanisms to break their topological symmetry. To study a possible
mechanism leading to symmetry breaking it seems natural to study  couplings of
topological Yang-Mills theory to  matter multiplets. It is in this context
where, indeed, a breaking of the topological symmetry could appear. The matter
models presented in this work are topological models. However, when the
coupling of these models to topological Yang-Mills is carried out such a
property is lost. It turns out that the  observables, although share many of
the properties as the ones in topological quantum field theories, acquire a
dependence on the metric of the four dimensional manifold. These results were
briefly reported in [\bts]. In this work we present a full account of the
results reported in [\bts] in what regards the structure of one particular
representation of the $N=2$ hypermultiplet [\fayet,\sohn], and we present the
study of another representation.

The second representation of $N=2$ supersymmetric matter treated in this paper
is based on the relaxed hypermultiplet [\howe]. Only a truncated version of
the theory resulting after the twisting is presented. The model is coupled to
topological Yang-Mills and the theory constructed in this way turns out to be
a topological quantum field theory. The observables are the same one as the
ones in topological Yang-Mills but in this case the observables acquire
corrections due to the presence of matter fields. The relation between the two
forms of topological matter in four dimensions is discussed in sect. 8.

Let us make a brief summary of how the paper is organized. In sect. 2, the
twisting of $N=2$ supersymmetry in four dimensions is performed obtaining the
resulting four dimensional topological algebra. It is shown that the twist is
unique up to reversal of orientation. The corresponding topological superspace
is constructed. In sect. 3, topological Yang-Mills is constructed in the
framework of topological superspace and its uniqueness is discussed. In sect.
4, a central extension of the topological algebra is presented, which is needed
since one of the representations of the hypermultiplet chosen in this work
possesses a non-vanishing central charge. The rest of the section deals with
the construction of the topological matter multiplet associated to the twisted
form of the representation of the $N=2$ hypermultiplet built in
[\fayet,\sohn]. In sect. 5, the coupling of the resulting topological matter
to topological Yang-Mills is carried out. Section 6 presents a truncated
version of the model presented in sect. 5. In sect. 7  the energy-momentum
tensors of the models considered in the previous sections are constructed and
analyzed. Section 8 presents a topological matter model related to a twisted
version of the relaxed hypermultiplet.  Finally, in sect. 9 we state our final
comments and remarks. An appendix describes the conventions used in this paper.

\endpage
\chapter{Topological Algebra in 4D}


In this section we will analyze the possible twistings of $N=2$ supersymmetry
in four dimensions. Our conclusion is that the twisting procedure is unique (up
to orientation reversal).

We will construct the $4D$ topological algebra by twisting the algebra of $N=2$
supersymmetry. This is the four-dimensional analogue of the construction
presented in  [\pablo]. As we will argue the twisting procedure is essentially
unique. Our starting point is the algebra of $N=2$ supersymmetry. We will
denote the Poincar\'e generators by $ P_{\alpha\dot\beta},   J_{\alpha\beta},
J_{\dot\alpha\dot\beta}$. For a summary of the index-convention used in this
paper see the appendix. Supersymmetry generators are denoted by $ Q_{a \alpha},
\overline{ Q}^{\,\,a}_{\dot\alpha}$ while internal $SU(2)$ generators by $
T_a^{\,\,b}$. The $N=2$ supersymmetry algebra takes the form \REF\book{J.
Gates, M. Grisaru, M. Ro\v cek and W. Siegel, ``Superspace" (Benjamin, 1984)}
[\book]:
$$
\eqalign{ \{Q_{a \alpha},  \overline{ Q}^{\,\,b}_{\dot\beta} \}&=
\delta_{a}^{\,\,b} P_{\alpha\dot\beta}, \cr \{Q_{a \alpha},  Q_{b \beta} \}&=
0,
\cr [Q_{a \alpha},  P_{\beta\dot\beta}]&= [ J_{\dot\alpha\dot \beta},  Q_{c
\gamma}]=0, \cr [ J_{\alpha\beta}, Q_{c \gamma}]& = { i\over2}
C_{\gamma(\alpha} Q_{c \beta)}, \cr [ J_{\dot\alpha\dot\beta},
\overline{Q}^{\,\,c}_{\dot\gamma}]& = { i\over2} C_{\dot\gamma(\dot\alpha}
\overline{Q}^{\,\,c}_{\dot \beta)}, \cr [ J_{\alpha\beta},
P_{\gamma\dot\gamma}]&= { i\over2}C_{\gamma(\alpha} P_{\beta)\dot\gamma}, \cr [
J_{\dot\alpha\dot\beta}, P_{\gamma\dot\gamma}]&= { i\over2}C_{\dot\gamma
(\dot\alpha} P_{\gamma\dot\beta) }. \cr} \qquad \eqalign{ [ J_{\alpha \beta},
J^{\gamma \delta}] &= -{ i\over2}\delta_{(\alpha}^{\,\,(\gamma}
J_{\beta)}^{\,\,\delta)}, \cr [ J_{\dot\alpha \dot\beta},  J^{\dot\gamma
\dot\delta}] &= -{ i\over2}\delta_{(\dot\alpha}^{\,\,(\dot\gamma}
J_{\dot\beta)}^{\,\,\dot\delta)}, \cr [ J_{\alpha \beta},  J^{\dot\gamma
\dot\delta}] &= [P_{\alpha\dot\alpha}, P_{\beta\dot\beta}]=0, \cr [
T_a^{\,\,b}, Q_{c \gamma}] &= -\half (\delta_c^{\,\,b} Q_{a \gamma} -	\half
\delta_a^{\,\,b} Q_{c \gamma}), \cr [ T_a^{\,\,b}, \overline{Q}^{\,\,c}_{\dot
\gamma}] &= \half (\delta_a^{\,\,c} \overline{Q}^{\,\,b}_{\dot \gamma} -\half
\delta_a^{\,\,b} \overline{Q}^{\,\,c}_{\dot \gamma}), \cr [ T_a^{\,\,b},
T_c^{\,\,d}] &= \half(\delta_a^{\,\,d}T_c^{\,\,b} -\delta_c^{\,\,b}
T_a^{\,\,d}),
\cr}  \eqn\uno
$$
 All other (anti)commutators vanish or are found by hermitian
conjugation. Notice that we do not consider central charges. The Lorentz
generator is symmetric in its two indices. The $SU(2)$ generators  $
T_a^{\,\,b}$ satisfy $ T_a^{\,\,a}=0$, being only three of them independent.
To carry out the twisting procedure it is convenient to introduce a matrix $
C_{ab}$ and its inverse $ C^{ab}$ to raise and lower isospin indices. These
matrices are antisymmetric and satisfy,
$$
 C_{ab}  C^{cd} = \delta_a^{\,\,c}\delta_b^{\,\,d}-\delta_a^{\,\,d}
\delta_b^{\,\,c}.
 \eqn\pera
 $$
 It allows to redefine the $SU(2)$ generators $
T_a^{\,\,b}$ in the more convenient form,
 $$
 T_{ab}= T_a^{\,\,c} C_{cb},
 \eqn\perita
$$ which turn out to be symmetric due
to the condition $ T_a^{\,\,a}=0$. In terms of the new $SU(2)$ generators the
entries of the algebra \uno\ which are modified by this redefinition become
$$
\eqalign{ [ T_{ab}, Q_{c \gamma}] &= -{1\over4}
 C_{c(b} Q_{a) \gamma}, \cr [ T_{ab}, \overline{ Q}^{\,\,c}_{\dot \gamma}] &=
{1\over4}\delta_{(a|}^{\,\,c}\overline{Q}_{b)\dot\gamma}, \cr
 [ T_{ab}, T_{cd}] &= -{1\over 4} C_{(c|(b}T_{a)|d)}.\cr } \eqn\dos $$ The
$N=2$ supersymmetry algebra \uno\ possesses an additional $U(1)$ symmetry.
This symmetry, whose generator will be denoted by $U$, is such that,
$$
[U,Q_{a\alpha}] = Q_{a\alpha}, \,\,\,\,\,\,\,\,\,\,\,\,\,\,\,\,\,\,\,\,
[U,\overline Q_{\dot\alpha}^a] = -\overline Q_{\dot\alpha}^a, \eqn\rolle
$$
while it acts trivially on the rest on the generators. This symmetry will turn
out to be the ghost number symmetry of the twisted theory.

The twisting procedure consists of a redefinition of the Lorentz generators and
an identification of the isospin indices as right-handed spin indices relative
to the new Lorentz generator:
$$
\tilde{ J}_{ AB}= J_{AB}-2iT_{AB}, \eqn\tres
$$
where capital letters will denote for the moment new spin indices. The
coefficient in \tres\ is uniquely determined by the requirement that $\tilde
J_{\alpha\beta}$ possess a commutator with itself as the one of a Lorentz
generator. The choice $J_{AB} \rightarrow  J_{\alpha\beta}$  is, however, a
matter of convention. The opposite choice (\ie, $J_{AB} \rightarrow
J_{\dot\alpha\dot\beta}$) would lead to a mirror image algebra (left-handed
$\leftrightarrow$ right-handed) of the one we are about to construct. If we
denote the  $SU(2)$ group associated to the generator $J_{\alpha\beta}$
($J_{\dot\alpha\dot\beta}$) as $SU(2)_\ele$  ($SU(2)_\ere$), and the internal
$SU(2)$ group as $SU(2)_\lai$, what we are doing in \tres\ is to replace
$SU(2)_\ele \times SU(2)_\ere$ by $SU(2)_\ele \times SU(2)_{\ere}'$, where
$SU(2)_{\ere}'$ is the diagonal sum of $SU(2)_\ere$ and $SU(2)_\lai$. The
choice opposite to the one taken in \tres\ would have led to $SU(2)_{\ele}'
\times SU(2)_{\ere}$. Using \tres\ one finds that the following combination of
components of the supersymmetry generator transforms as a scalar under
$\tilde{ J}_{ AB}$:
$$ [\tilde{ J}_{ AB}, \quad Q_{+-}-  Q_{-+}] = 0,
\eqn\cuatro
$$
which leads to the definition of the ``scalar" under the new
Lorentz transformations: $$
 Q= -i ( Q_{+-}-  Q_{-+}) =  C_{AB} Q^{AB}.
 \eqn\cinco
$$
Furthermore, one finds that $$ \{ Q,  Q\} =0,  \eqn\seis $$
which gives a first indication of the topological structure of the resulting
algebra. The choice \cinco\ is {\sl unique} up to a constant, \ie, \cinco\ is
the unique linear combination of components of $ Q_{a \alpha}$ (up to a global
factor) such that it behaves as a scalar under the new Lorentz generators
\tres\ and satisfies $ Q^2 =0$. The rest of the components of $ Q_{a \alpha}$
build a symmetric generator, $$
 H_{AB} = Q_{(AB)}.  \eqn\siete $$ Finally, it is straightforward to show that
the rest of the SUSY generators, $\overline{ Q}^{\,\,a}_{\dot\alpha}$, build a
generator, $$
 G_{A\dot A} =  C_{BA} \overline{Q}^{\,\,B}_{\dot A},  \eqn\ocho $$ which
transforms as a vector under the new Lorentz generator \tres: $$ [\tilde{
J}_{AB}, G_{ C\dot C}] = { i \over 2} C_{C(A}G_{B)\dot C}. \eqn\nueve $$ It is
now simple to work out the full form of the topological algebra in terms of
its defining generators $ Q, H_{\alpha\beta}, G_{\alpha \dot\alpha},
J_{\alpha\beta}, J_{\dot\alpha \dot\beta}, P_{\alpha \dot\beta}$ where we have
dropped the tilde from $\tilde{ J}_{AB}$ and renamed the spin indices with
capital letters by the standard Greek notation. We underline commuting vector
indices:
$$
\eqalign{ \{Q,Q\}&=0, \cr
 \{ Q,H_{\alpha\beta}\}&=0, \cr \{ H_{\alpha\beta},H_{\gamma\delta}\}&=0, \cr\{
Q,G_{\alpha\dot\beta}\}&= P_{\underline{\alpha\dot\beta}}, \cr  \{
H_{\alpha\beta},G_{\gamma\dot\delta}\}&=  C_{(\alpha|\gamma|}
 P_{\underline{\beta)\dot\delta}}, \cr
\{G_{\alpha\dot\beta},G_{\gamma\dot\delta}\}&=0, \cr [
Q,P_{\underline{\alpha\dot\beta}}] &=
[H_{\gamma\delta},P_{\underline{\alpha\dot\beta}}] =0, \cr
[P_{\underline{\alpha\dot\beta}},P_{\underline{\gamma\dot\delta}}]&=
[G_{\gamma\dot\delta},P_{\underline{\alpha\dot\beta}}]=0, \cr [Q,\tilde{
J}_{\alpha\beta}]&=[ Q,\tilde{ J}_{\dot\alpha\dot\beta}]=0. \cr} \qquad
\eqalign{[\tilde{ J}_{\alpha\beta},H_{\gamma\delta}]&= {i\over2}
C_{(\gamma|(\alpha} H_{\beta)|\delta)}, \cr
 [\tilde{ J}_{\alpha\beta},G_{\gamma\dot\delta}]&= {i\over2} C_{\gamma(\alpha}
G_{\beta)\dot\delta}, \cr [\tilde{ J}_{\alpha\beta},\tilde{ J}_{\gamma\delta}]&
=  -{i\over2} C_{(\alpha|(\gamma}\tilde{ J}_{\delta)|\beta)},  \cr
[H_{\gamma\delta},\tilde{ J}_{\dot\alpha\dot\beta}]&=0, \cr [\tilde{
J}_{\dot\alpha\dot\beta},G_{\gamma\dot\delta}]&= {i\over2}
C_{\dot\delta(\dot\alpha} G_{\gamma\dot\beta)}, \cr [\tilde{
J}_{\alpha\beta},\tilde{ J}_{\dot\gamma\dot\delta}] &= 0, \cr [\tilde{
J}_{\dot\alpha\dot\beta},\tilde{ J}_{\dot\gamma\dot\delta}] &=
 {i\over 2} C_{(\dot\gamma|(\dot\beta}\tilde{ J}_{\dot\alpha)|\dot\delta)},
\cr } \eqn\diez$$

The essential feature of this algebra, which encodes its topological character,
is contained in the anticommutator between $Q$ and $ G_{\alpha\dot\beta}$. This
anticommutator expresses that the translation generator $ P_{\alpha\dot\beta}$
is $Q$-exact. This is a necessary condition for a theory to have an
energy-momentum tensor which is $Q$-exact and therefore topological. Note that
a nilpotent self-dual operator $ H_{\alpha\beta}$ is present in this algebra.
The mirror image algebra, which would have resulted from the choice $ J_{AB}
\rightarrow
 J_{\dot\alpha \dot\beta}$ in \tres\ would have contained an anti-self-dual
operator $ H_{\dot\alpha \dot\beta}$.

{}From \rolle, \cinco, \siete\ and \ocho, it turns out that the $U(1)$ charges
of
the new generators are, $$ [U,Q] = Q, \,\,\,\,\,\,\,\,\,\,\,
[U,H_{\alpha\beta}] = H_{\alpha\beta}, \,\,\,\,\,\,\,\,\,\,\,
[U,G_{\alpha\dot\beta}]= - G_{\alpha\dot\beta}, \eqn\picard $$ while it is
zero for the rest of the generators. In the twisted algebra the charges in
\picard\ are called ghost numbers.

Our next task is to construct the superspace corresponding to the topological
algebra \diez. Besides the space-time coordinates
$x^{\underline{\alpha\dot\beta}}$ we introduce anticommuting coordinates
$\theta$, $\theta^{\alpha\beta}$ and $\theta^{\alpha\dot\beta}$, associated to
the odd generators $Q$, $H_{\alpha\beta}$ and $G_{\alpha\dot\beta}$,
respectively. A point in superspace is therefore labeled by 4+8 quantities
$x^{\underline{\alpha\dot\beta}}$, $\theta$, $\theta^{\alpha\beta}$ and
$\theta^{\alpha\dot\beta}$. The representation of the operators entering
\diez\ in terms of these coordinates is the following: $$ \eqalign{
P_{\underline{\alpha\dot\beta}} & = i {\partial\over \partial
x^{\underline{\alpha\dot\beta}}}, \cr Q & ={\partial\over \partial \theta} + {i
\over 2} \theta^{\alpha\dot\beta} {\partial\over \partial
x^{\underline{\alpha\dot\beta}}}, \cr  H_{\alpha\beta} & = {\partial\over
\partial \theta^{\alpha\beta} } +
 {i \over 2} C_{(\alpha|\gamma}\theta^{\gamma\dot\beta} {\partial\over \partial
x^{\underline{\beta)\dot\beta}}}, \cr G_{\alpha\dot\beta} & = {\partial\over
\partial  \theta^{\alpha\dot\beta} } +
 {i \over 2} \theta {\partial\over \partial x^{\underline{\alpha\dot\beta}}} -
{i\over2}C_{\alpha\delta} \theta^{\gamma\delta}{\partial\over \partial
x^{\underline{\gamma\dot\beta}}}.\cr} \eqn\limon $$

Superspace covariant derivatives $D$, $D_{\alpha\beta}$ and
$D_{\alpha\dot\beta}$ are introduced as operators which (anti)commute with
$P_{\alpha\dot\beta}$, $Q$, $H_{\alpha\beta}$ and $G_{\alpha\dot\beta}$. Their
representation in terms of the superspace coordinates is, $$ \eqalign{ D &
=i\Big({\partial\over \partial \theta} - {i \over 2} \theta^{\alpha\dot\beta}
{\partial\over \partial x^{\underline{\alpha\dot\beta}}}\Big), \cr
D_{\alpha\beta} & = i\Big({\partial\over \partial  \theta^{\alpha\beta} } -
 {i \over 2} C_{(\alpha|\gamma}\theta^{\gamma\dot\beta} {\partial\over \partial
x^{\underline{\beta)\dot\beta}}}\Big), \cr D_{\alpha\dot\beta} & =
i\Big({\partial\over \partial \theta^{\alpha\dot\beta} } -
 {i \over 2} \theta {\partial\over \partial x^{\underline{\alpha\dot\beta}}} +
{i\over2}C_{\alpha\delta} \theta^{\gamma\delta}{\partial\over \partial
x^{\underline{\gamma\dot\beta}}}\Big).\cr} \eqn\limonero $$ The prefactors are
chosen for later convenience. It is simple to verify that, indeed, $$ \eqalign{
\{D,Q\} & = \{D,H_{\alpha\beta}\} = \{D,G_{\alpha\dot\beta}\} = 0, \cr
\{D_{\alpha\beta},Q\} & = \{D_{\alpha\beta},H_{\gamma\delta}\} =
\{D_{\alpha\beta},G_{\gamma\dot\delta}\} = 0, \cr \{D_{\alpha\dot\beta},Q\} &
= \{D_{\alpha\dot\beta},H_{\gamma\delta}\} =
\{D_{\alpha\dot\beta},G_{\gamma\dot\delta}\} = 0. \cr} \eqn\naranja $$ The
algebra of the superspace covariant derivatives turns out to be the following,
$$ \eqalign{ \{D,D\} & = \{D_{\alpha\beta},D_{\gamma\delta}\} =
\{D,D_{\alpha\beta}\}= \{D_{\alpha\dot\beta},D_{\gamma\dot\delta}\}= 0, \cr
\{D,\partial_{\underline{\alpha\dot\beta}}\} & =
\{D_{\alpha\beta},\partial_{\underline{\gamma\dot\delta}}\}=
\{D_{\alpha\dot\beta},\partial_{\underline{\gamma\dot\delta}}\}=0,\cr
\{D,D_{\alpha\dot\beta}\} & = i\partial_{\underline{\alpha\dot\beta}}, \cr
\{D_{\alpha\beta},D_{\gamma\dot\delta}\} & =  i
C_{(\alpha|\gamma}\partial_{\underline{\beta)\dot\delta}}. \cr} \eqn\banana $$

The superspace formulation allows to construct theories which contain all the
symmetries present in the topological algebra \diez. The procedure is first to
introduce multiplets and then suitable actions which fix their kinematics and
interactions. Matter multiplets are usually constructed  by imposing superspace
covariant constraints on tensor superfields. These constraints involve the
superspace covariant derivatives \limonero, and therefore due to the relations
\naranja\ their covariance under the topological algebra is guaranteed. Other
multiplets as, for example, vector multiplets, are introduced by constructing
gauge superspace covariant derivatives and then imposing gauge covariant
constraints on their algebra. In the next section we will present in this
framework a theory for the vector multiplet which is equivalent to topological
Yang-Mills theory as formulated in [\tqft], and we will study other possible
sets of constraints which might lead to other models.

\endpage
\chapter{Topological vector multiplet}

Let us consider a gauge group $G$ and connections $A$, $A_{\alpha\beta}$,
$A_{\alpha\dot\beta}$ and $A_{\underline{\alpha\dot\beta}}$ which allow to
define gauge superspace covariant derivatives: $$ \eqalign{ \D & =D-iA, \cr
\D_{\alpha\beta} & =D_{\alpha\beta}-iA_{\alpha\beta}, \cr \D_{\alpha\dot\beta}
& =D_{\alpha\dot\beta}-iA_{\alpha\dot\beta}, \cr
\D_{\underline{\alpha\dot\beta}} &
=D_{\underline{\alpha\dot\beta}}-iA_{\underline{\alpha\dot\beta}}. \cr}
\eqn\fresa $$ Notice that $A$, $A_{\alpha\beta}$ and $A_{\alpha\dot\beta}$ are
anticommuting superfields. Often we will use a condensed notation for
superspace indices. We will denote by $I$ the set of indices
$_{-},\alpha\beta,\alpha\dot\beta,\underline{\alpha\dot\beta}$, by $J$ the set
$_{-},\gamma\delta,\gamma\dot\delta,\underline{\gamma\dot\delta}$, etc. Notice
that $\_$ refers to no index, \ie, $\D\equiv\D_{-}$ or $A\equiv A_{-}$. Using
this notation \fresa\ takes the condensed form, $$ \D_{I}=D_{I}-iA_{I}.
\eqn\doce $$ These gauge superspace covariant derivatives possess the
following superspace algebra, $$ [\D_{I},\D_{J}\} =T_{IJ}{}^{K}\D_{K} -iF_{IJ},
\eqn\kiwi $$ where $T_{IJ}{}^{K}$ is the superspace torsion in \banana, $$
\eqalign{ {T_{-,\alpha\dot\beta}}{}^{\underline{\gamma\dot\delta}}
&=i\delta_{\alpha}^{\,\,{\gamma}} \delta_{\dot\beta}^{\,\,\,\,{\dot\delta}},
\cr {T_{\alpha\beta,
\gamma\dot\delta}}{}^{\underline{\rho\dot\sigma}}&=iC_{(\alpha|\gamma}
\delta_{\beta)}^{\,\,\,\,{\rho}} \delta_{\dot\delta}^{\,\,\,\,{\dot\sigma}},
\cr } \eqn\trece $$ while all other components are zero. In \kiwi, $F_{IJ}$
are superfield strengths. Constraints on the form of these superfield
strengths define the vector multiplet. The constraints leading to
Donaldson-Witten theory  are $$ \eqalign{ F_{-,-}&=\half V,  \cr
F_{-,\alpha\beta}&=0, \cr F_{\alpha\beta,\gamma\delta}&=\half
C_{\alpha(\gamma|}C_{\beta|\delta)} V, \cr F_{-,\gamma\dot\delta}&=0, \cr
F_{\alpha\beta,\gamma\dot\delta}&=0, \cr
F_{\alpha\dot\beta,\gamma\dot\delta}&=C_{\alpha\gamma}C_{\dot\beta\dot\delta}W,
\cr } \eqn\catorce $$ where $V$ and $ W$ are scalar superfields. From these
constraints and \picard\ follow that the superfields $V$ and $W$ have ghost
numbers 2 and -2 respectively: $$ [U,V] = 2 V, \,\,\,\,\,\,\,\,\,\,\,\,\,\,\,
[U,W] = -2 W. \eqn\hilbert $$

Once the constraints \catorce\ are taken into account the Bianchi identities
satisfied by the gauge superspace covariant derivatives \fresa\ provide
relations among the  scalar superfields $V$ and $ W$ and the gauge
superconnections. The analysis of these identities is long and tedious and we
just list here its outcome. The resulting relations can be classified in three
types. There are linear constraints on the superfields  $V$ and $ W$, $$ \D  V
=0, \,\,\,\,\, \D_{\alpha\beta} V=0, \,\,\,\,\, \D_{\alpha\dot\beta}W=0,
\eqn\lima $$ expressions for the superfield strengths which are not in
\catorce, $$ \eqalign{ F_{\_,\underline{\alpha\dot\beta}}&=
{i\over4}\D_{\alpha\dot\beta} V, \cr
F_{\gamma\delta,\underline{\alpha\dot\beta}}&=
-{i\over4}C_{(\gamma|\alpha}\D_{\delta)\dot\beta} V, \cr
F_{\alpha\dot\beta,\underline{\gamma\dot\delta}}&=
{i\over2}C_{\dot\beta\dot\delta}\big[C_{\alpha\gamma}\D W + \D_{\alpha\gamma}
W \big],\cr F_{\underline{\alpha\dot\beta},\underline{\gamma\dot\delta}}&=
\half C_{\dot\beta\dot\delta}\D\D_{\alpha\gamma}W +
{1\over8}\big[\D_{\alpha\dot\beta}, \D_{\gamma\dot\delta}\big] V, \cr }
\eqn\quince $$ and, finally, second order constraints among the fields $V$ and
$W$, $$ \D_{\gamma\delta}\D_{\sigma\alpha}W= \half
C_{(\gamma|(\sigma}C^{\dot\beta\dot\tau}F_{\underline{\alpha)\dot\beta},
\underline{|\delta)\dot\tau}}
-{1\over8}C_{(\gamma|(\alpha}
\D_{\sigma)}^{\,\,\,\,\dot\beta}\D_{|\delta)\dot\beta}V.
\eqn\quincebis
$$
The relations \lima, \quince\ and \quincebis, are very
important in the analysis of the independent component fields of the theory.
They represent the consequences of the relations \catorce\ which define the
vector-like topological multiplet.

The theory under construction possesses the full symmetry generated by the
topological algebra \diez. Of particular importance are the odd symmetries
generated by $Q$, $H_{\alpha\beta}$ and $G_{\alpha\dot\beta}$. Let $\Phi$ be a
generic superfield. The transformations corresponding to these symmetries take
the form, $$ \eqalign{\delta\Phi &=i\epsilon Q\Phi, \cr \delta'\Phi&=
i\epsilon^{\alpha\beta}H_{\alpha\beta}\Phi, \cr \delta''\Phi&=
i\epsilon^{\alpha\dot\beta} G_{\alpha\dot\beta}\Phi, \cr} \eqn\sandia $$ where
$\epsilon$, $\epsilon^{\alpha\beta}$ and $\epsilon^{\alpha\dot\beta}$ are
scalar, self-dual and vector constant anticommuting parameters. The theory is
also invariant under gauge transformations. These take the form, $$
\tilde\delta_K  A_I = D_I K, \eqn\watermelon $$ where $A_I$ is any of the
connections in \fresa\ and $K$ an arbitrary scalar superfield.

Superspace actions are defined as integrations over superspace. The full
superspace measure has the 8 anticommuting coordinates $\theta$,
$\theta^{\alpha\beta}$ and $\theta^{\alpha\dot\beta}$ plus the ordinary 4
space-time commuting coordinates. The dimension of this measure is therefore 0
if one associate the standard value 1/2 to the dimensions of $\theta$,
$\theta^{\alpha\beta}$ and $\theta^{\alpha\dot\beta}$. This implies that it is
not possible to write a suitable action unless one takes only part of the
superspace measure [\book]. In the untwisted analysis this would correspond to
the choice of a chiral measure. In virtue  of constraints \lima\ there exist an
essentially {\sl unique} suitable action with zero ghost number which involves
 the fields
$V$ and $W$ and is $Q$-exact. This action is: $$ S_0= \int d^4x d^4\theta \tr(
W^2 ), \eqn\coco $$ where $d^4\theta$ denotes the measure built from $\theta$
and  $\theta^{\alpha\beta}$: $d\theta d\theta_{11}d\theta_{12} d\theta_{22}$.
The presence of $d\theta$ in the measure assures that the action is $Q$-exact.
Another suitable superspace action of zero ghost number can be built making
use of the part of the full superspace measure not present in \coco:
$\widehat{d^4\theta}$: $d\theta_{1\dot 1}d\theta_{1\dot 2} d\theta_{2\dot 1}
d\theta_{2\dot 2}$. Taking into account \lima\ the only non-trivial choice is:
$$ S_1= \int d^4x \widehat{d^4\theta} \tr( V^2 ). \eqn\cocouno $$ Both, $S_0$
and $S_1$, are invariant under the symmetries generated by the generators
\diez, and turn out to be equivalent. Their Lagrangians differ by a total
derivative which is proportional to the integrand of the second Chern class.

Our next task is to formulate the theory in terms of component fields. We will
do this projection in a covariant  approach taking a Wess-Zumino gauge [\book].
The form of the supergauge transformation \watermelon\ indicates that the odd
connections, say $A_{\alpha\beta}$, transform as, $$ \tilde\delta_K
A_{\alpha\beta} = {\partial\over \partial\theta^{\alpha\beta}} K + ...
\eqn\apple $$ and therefore the lowest component of the superfield
$A_{\alpha\beta}$ can be gauged away algebraicaly using one of the higher
components of $K$. The Wess-Zumino gauge consists of making a gauge choice
while projecting into component fields of the type that we have just described
for all the components of the odd connections and the higher components of the
even connection $A_{\underline{\alpha\dot\beta}}$. Some of these components
are expressed via \catorce\ and \quince\ in terms of the components of the
fields $V$ and $W$.  One is left with only the lowest component of the even
connection $A_{\underline{\alpha\dot\beta}}$, the gauge invariance
corresponding to the lowest component of $K$, and the components of the fields
$V$ and $W$ which are independent after taking into account the constraints
\lima, \quince\ and \quincebis. We define these independent components as,
$$
\eqalign{W|&=2^{1/2}\lambda, \cr \D W| &=2^{5/4}\eta, \cr \D_{\alpha\beta} W|
&=2^{5/4}\chi_{\alpha\beta}, \cr
 V|&=2^{3/2}\phi, \cr \D_{\alpha\dot\beta}  V| &=
-i2^{3/4}\psi_{\alpha\dot\beta}, \cr
\D_{(\alpha\dot\sigma}\D_{\beta)}^{\,\,\dot\sigma}  V| &=
8G_{\alpha\beta}, \cr
} \eqn\berry
$$
The numerical factors introduced in \berry\ are such that the
final form of the theory coincides with the one in [\tqft]. In \berry\ $|$
means theta-independent part. Notice that our projection procedure is
covariant. All component fields appearing in \berry\ transform in the adjoint
representation under gauge transformations with parameter $\kappa=K|$. On the
other hand, the only component of the connections left,
$A_{\underline{\alpha\dot\beta}}|$, which will be denoted simply as
$A_{\underline{\alpha\dot\beta}}$ transforms in the standard way under gauge
transformations, $$ \tilde\delta_\kappa A_{\underline{\alpha\dot\beta}} =
\D_{\underline{\alpha\dot\beta}}\kappa. \eqn\trufa $$

Our next task is to compute the symmetry transformations of the component
fields generated by $Q$, $H_{\alpha\beta}$ and $G_{\alpha\dot\beta}$. To carry
this out one must project the transformations \sandia\ taking into account that
the Wess-Zumino gauge has been chosen. Let us compute as an example of the
procedure the transformation of the field $\eta$ in \berry\ under the symmetry
generated by $Q$. One finds, $$ \delta \eta = i\epsilon Q \D W |= \epsilon \D
\D W |, \eqn\ice $$ where in the last step the fact that we work in a
Wess-Zumino gauge has been used. The first of the constraints \catorce\ allows
to write \ice\ as, $$ \delta \eta =  -{i\over 4} 2^{-5/4}\epsilon [V,W]| = i
2^{-5/4}\epsilon [\lambda,\phi] \equiv \epsilon'{i\over 2}[\lambda, \phi],
\eqn\cream $$ where in the last step a redefinitions of $\epsilon$ has been
utilized. This example illustrates the general procedure. After replacing
$\epsilon' \rightarrow \epsilon$, the form of the symmetry transformation
generated by $Q$ turns out to be:
$$
\eqalign{ \delta \lambda &= 2\epsilon\eta,
\cr \delta \eta &= {i\over 2}\epsilon[\lambda, \phi], \cr \delta \phi &= 0,\cr
\delta \chi_{\alpha\beta} &= \epsilon\half(F^{+}_{\alpha\beta} -
G_{\alpha\beta}), \cr \delta \psi_{\alpha\dot\beta} &=-\epsilon
\D_{\underline{\alpha\dot\beta}} \phi, \cr \delta G_{\alpha\beta}&=
\epsilon\big(\D_{\underline{(\alpha\dot\sigma}}\psi_{\beta)}^{\,\,\dot\sigma}
-2i\big[\chi_{\alpha\beta}, \phi\big]\big), \cr   \delta
A_{\underline{\alpha\dot\beta}} &=\epsilon \psi_{\alpha\dot\beta}, \cr }
\eqn\diecisiete
$$
where $F^{\pm}_{\alpha\beta}$ represents the self-dual and
anti-self-dual parts of the gauge field strength, $$
 F^{+}_{\alpha\beta}= C^{\dot\beta\dot\delta}F_{\underline{\alpha\dot\beta},
\underline{\gamma\dot\delta}}, \,\,\,\,\,\,\,\,\,\,\,\,
 F^{-}_{\dot\alpha\dot\beta}= C^{\sigma\rho}F_{\underline{\sigma\dot\alpha},
\underline{\rho\dot\beta}}. \eqn\dieciseis $$ The transformation of the
component fields under the rest of the symmetries in \sandia\ take the
following form: $$\eqalign{ \delta'\lambda &=
2\epsilon^{\alpha\beta}\chi_{\alpha\beta},\cr   \delta' \eta &=\half
\epsilon^{\alpha\beta}(G_{\alpha\beta}-F^{+}_{\alpha\beta}),\cr \delta' \phi &=
0, \cr \delta'\chi_{\alpha\beta} &= \half
\epsilon^{\gamma\delta}\big\{C_{\gamma(\alpha}\big[F^{+}_{\beta)\delta}
+G_{\beta)\delta}\big]-iC_{\gamma(\alpha}C_{\beta)\delta}\big[\lambda,\phi\big]
\big\},\cr \delta' \psi_{\alpha\dot\beta}
&=2\epsilon^{\gamma\delta}C_{\alpha\gamma}\D_{\underline{\delta\dot\beta}}\phi,
\cr  \delta'G_{\alpha\beta}&=2\epsilon^{\gamma\delta}C_{\gamma(\alpha}
\big(\D_{\underline{\delta\dot\sigma}}\psi_{|\beta)}^{\,\,\dot\sigma}-
i\big[\eta C_{|\beta)\delta} +\chi_{|\beta)\delta},\phi\big]\big), \cr
\delta'A_{\underline{\alpha\dot\beta}}&=2\epsilon^{\gamma\delta}
C_{\alpha\gamma}\psi_{\delta\dot\beta}, \cr} \eqn\dieciocho
$$
$$\eqalign{
\delta{''}\lambda &=0, \cr
\delta{''}\eta&={i\over2}\epsilon^{\alpha\dot\beta}\D_{\underline
{\alpha\dot\beta}}\lambda,\cr \delta{''}\chi_{\alpha\beta}&=
{i\over2}\epsilon^{\gamma\dot\delta}C_{(\alpha|\gamma},
\D_{\underline{\beta)\dot\delta}}\lambda
\cr \delta{''}\phi&=-i\epsilon^{\alpha\dot\beta}\psi_{\alpha\dot\beta}, \cr
\delta{''}\psi_{\alpha\dot\beta}&={i\over2}\epsilon^{\gamma\dot\delta}
\big\{C_{\gamma\alpha}
F^{-}_{\dot\beta\dot\delta} +
C_{\dot\beta\dot\delta}\big(iC_{\gamma\alpha}\big[\phi,\lambda \big] +
G_{\alpha\gamma}\big)\big\}, \cr \delta''G_{\alpha\beta}&=
\epsilon^{\sigma\dot\tau}\big\{iC_{(\beta|\sigma}
\D_{\underline{\alpha)\dot\tau}}\eta
+C_{(\alpha|\sigma} \big[\psi_{\beta)\dot\tau},\lambda \big]
+i\D_{\underline{(\alpha\dot\tau}}\chi_{\beta)\sigma}-2i
\D_{\underline{\sigma\dot\tau}}\chi_{\alpha\beta} \big\}, \cr
\delta''A_{\underline{\alpha\dot\beta}}&=
i\epsilon^{\gamma\dot\delta}C_{\dot\delta\dot\beta}\big(C_{\gamma\alpha}\eta +
\chi_{\alpha\gamma}\big). \cr } \eqn\diecinueve
$$
These symmetries close among
themselves as dictated by \diez\ up to gauge transformations. Finally, the $U$
transformations of the component fields, or ghost numbers, can be computed
easily from \picard, \hilbert\ and \berry. They are, $$ \eqalign{ [U,\lambda]
=& -2 \lambda,\cr [U,\eta] = & - \eta,\cr} \qquad \eqalign{
[U,\chi_{\alpha\beta}] =& - \chi_{\alpha\beta},\cr [U,\phi] =& 2 \phi,\cr}
\qquad \eqalign{ [U,\psi_{\alpha\dot\beta}] =& \psi_{\alpha\dot\beta},\cr
[U,G_{\alpha\beta}] =& 0. \cr} \eqn\riemann $$ Of course, the gauge field
$A_{\underline{\alpha\dot\beta}}$ has zero ghost number.

The action $S_0$ in \coco\ can be easily written in terms of component fields
after taking into account that,
$$
\eqalign{ S_0= & {1\over e^2}\int d^4x
d^4\theta \tr( W^2 ) = {1\over 6e^2}\int d^4x D D_\alpha{}^\beta
D_\beta{}^\gamma D_\gamma{}^\alpha \tr( W^2 )| \cr = &
{1\over 6e^2}\int d^4x D \tr\D_\alpha{}^\beta \D_\beta{}^\gamma
\D_\gamma{}^\alpha ( W^2 ) |, \cr}
\eqn\pertiga
$$
where again use has been made of the fact that a Wess-Zumino gauge has been
chosen. In \pertiga\ $e$ is a dimensionless gauge coupling constant. Using the
constraints of the theory and the definitions \berry\ one finds, up to
irrelevant global factors  $$\eqalign{ S_0 = &{1\over e^2} \int d^4x \tr\Big\{
Q, \quad \half\big(F_{\alpha\gamma}^{+}+
G_{\alpha\gamma}\big)\chi^{\alpha\gamma}
+2\lambda\D_{\underline{\alpha\dot\tau}}\psi^{\alpha\dot\tau}
+2i\lambda\big[\eta,\phi\big]\Big\} \cr = & {1\over e^2} \int d^4x \tr
\Big\{{1\over8}\big(F_{+}^2 -G^2\big)  -
\chi^{\alpha\gamma}\D_{\underline{\alpha\dot\beta}}\psi_{\gamma}^{\,\,\dot\beta}
-{i\over2}\phi\big\{\chi_{\alpha\gamma},\chi^{\alpha\gamma}\big\}+
2\eta\D_{\underline{\alpha\dot\tau}}\psi^{\alpha\dot\tau} \cr
&-i\lambda\big\{\psi_{\alpha\dot\tau},\psi^{\alpha\dot\tau}\big\}
-\lambda\D_{\underline{\alpha\dot\tau}}\D^{\underline{\alpha\dot\tau}}\phi
+2i\phi\big\{\eta,\eta\big\}+{\big[\lambda,\phi\big]}^2 \Big\}. \cr}
\eqn\veintidos $$

This action is invariant under all the  symmetry transformations of the theory
\dieciseis, \dieciocho\ and \diecinueve, as well as under gauge
transformations. If one generalizes this action to an arbitrary curved four
dimensional manifold by introducing a metric tensor it turns out that it is
again {\sl invariant} under all symmetries provided the parameters of the first
three are covariantly constant: $$ {\cal D}_{\alpha\dot\beta} \epsilon =0,
\,\,\,\,\,\,\,\,\,\,\,\,\,\,\, {\cal D}_{\alpha\dot\beta}
\epsilon_{\gamma\sigma} =0, \,\,\,\,\,\,\,\,\,\,\,\,\,\,\, {\cal
D}_{\alpha\dot\beta} \epsilon_{\gamma\dot\sigma} =0, \eqn\turron $$ where
${\cal D}_{\alpha\dot\beta}$ is a covariant derivative which contains gauge
and Christoffel connections. Certainly, it is not guaranteed in general that a
covariantly constant vector and a self-dual tensor exist for an arbitrary
four-manifold. Thus, in general the $H_{\alpha\beta}$ and $G_{\alpha\dot\beta}$
symmetries do not exist. On the other hand, if one insists in having these
symmetries one is led to topological gravity. Namely, if these covariantly
constant vector and self-dual tensor do not exist let us gauge these global
symmetries so that the parameters become arbitrary. This is the form advocated
in [\pablo] to build couplings to topological gravity in two dimensions. A
similar construction should be carried out in four dimensions. It would be
interesting if this approach to four dimensional topological gravity
\REF\tgwi{E. Witten\journal\pl&B206(88)601} \REF\tgus{J.M.F. Labastida and M.
Pernici\journal\pl&B213(88)319} \REF\tgmp{R.C. Myers and V.
Periwal\journal\np&B333(90)536} \REF\tgmy{R.C. Myers\journal\pl&B252(90)365}
[\tgwi-\tgmy] leads to a theory of the type recently built in \REF\fre{D.
Anselmi and P. Fre\journal\np&392(93)401\journal\np&404(93)288} [\fre].

Let us  recall here  the form of the observables in Donaldson-Witten theory.
These are built out of the following basic ones [\tqft]. Let $\gamma$ be a
homology cycle on $M$, of dimension $k_\gamma$, and  $W_{k_\gamma}$ for
$k=0,1,...,4$ the differential forms, $$ \eqalign{ W_0 = & {1\over 2} \tr
\phi^2, \cr W_1 = & \tr(\phi\wedge\psi), \cr W_2 = & \tr({1\over
2}\psi\wedge\psi + i \phi\wedge F), \cr W_3 = & i\tr(\psi \wedge F),\cr W_4 =
& - {1\over 2} \tr (F\wedge F), \cr} \eqn\obs $$ then the basic observables of
the theory are $$ {\cal O}^{(\gamma)}=\int_\gamma W_{k_\gamma}.
\eqn\observables $$

The superspace approach considered in this paper constitutes an excellent
framework to build new topological quantum field theories involving gauge
fields. The constraints \catorce\ which define the theory could be properly
modified to obtain new theories. The constraints chosen in \catorce\ are a
particular choice but certainly one could imagine many others. We have tried
other possible sets of constraints and we have found that the ones in
\catorce\ are rather special. All our attempts to build new topological
quantum field theories modifying \catorce\ have led us either to trivial
theories, or theories involving fields of higher spin. We have not been able to
construct a non-trivial theory  with fields of spin no higher than two other
than the one leading to Donaldson-Witten theory. In this respect topological
Yang-Mills theory seems rather unique. These results, plus the fact that there
exist only one type of twisting, seem to indicate that type B models associated
to topological Yang-Mills do not exist unless one introduces additional field
content. In the rest of this paper we study matter couplings to
Donaldson-Witten theory.

\endpage
\chapter{Topological matter multiplets}

Our starting point will be the representation of the $N=2$ hypermultiplet
formulated in [\fayet,\sohn]. This multiplet has a non-vanishing central charge
and therefore, an extension of the topological algebra \diez\ is required.
 The final form of the extended algebra is obvious from the form of the
twisting. The only relations in \diez\ which change are, $$ \eqalign{ \{ Q,
Q\} &= Z, \cr \{H_{\alpha\beta},H_{\gamma\delta}\}&=
C_{\alpha(\gamma|}C_{\beta|\delta)}Z, \cr} \qquad \eqalign{
\{G_{\alpha\dot\beta},G_{\gamma\dot\delta}\}&=
C_{\alpha\gamma}C_{\dot\beta\dot\delta} Z, \cr [Z, \rm{anything} \} &= 0, \cr}
\eqn\calabaza $$ where $Z$ is the central charge generator. Notice that one
could still have generalized the extended topological algebra introducing a
dimensionless constant in the anticommutator
$\{H_{\alpha\beta},H_{\gamma\delta}\}$. We have analyzed this possibility and
it seems impossible to construct invariant actions unless such a dimensionless
constant is one. As we will describe below, the presence of central charges
makes the construction of invariant actions very restrictive. Notice also that
the relations \calabaza\ break the $U(1)$ symmetry. Taking into account
\picard, the best we can do is to maintain a $Z_4$ ghost number symmetry
assigning  $U$-charge 2 to the central charge generator $Z$. Indeed, with this
assignment, the $U(1)$ symmetry is preserved by the relations \calabaza\ modulo
4.

Let us introduce the extended superspace corresponding to \calabaza. Let $z$ be
a new real commuting coordinate corresponding to the central charge generator
$Z$. We define: $$ \eqalign{ Z &= i{\partial\over\partial z}, \cr Q &= Q^{(0)}
+{i\over2}\theta{\partial\over\partial z}, \cr H_{\alpha\beta} &=
H_{\alpha\beta}^{(0)} +{i\over2}\theta_{\alpha\beta}{\partial\over\partial
z},\cr G_{\alpha\dot\beta} &=G_{\alpha\dot\beta}^{(0)}
+{i\over2}\theta_{\alpha\dot\beta}{\partial\over\partial z},\cr} \eqn\pomelo
$$ where the superscript $(0)$ refers to the operators without central charge.
Superspace covariant derivatives are introduced as operators which
(anti)commute with $P_{\underline{\alpha\dot\beta}}$, $Q$, $H_{\alpha\beta}$,
and $G_{\alpha\dot\beta}$. Their representation in terms of superspace
coordinates is: $$ \eqalign{ D &= D^{(0)} +\half\theta{\partial\over\partial
z},\cr D_{\alpha\beta} &= D_{\alpha\beta}^{(0)}
+\half\theta_{\alpha\beta}{\partial\over\partial z}, \cr D_{\alpha\dot\beta} &=
D_{\alpha\dot\beta}^{(0)} +\half\theta_{\alpha\dot\beta}{\partial\over\partial
z}.\cr} \eqn\papaya $$ The factors are chosen for later convenience. They
verify the following anticommutation relations: $$ \eqalign{ \{D, D \} &=
i{\partial\over\partial z}, \cr \{D_{\alpha\beta},D_{\gamma\delta}\} &=
iC_{(\alpha|\gamma}C_{\beta)\delta}{\partial\over\partial z}, \cr
\{D_{\alpha\dot\beta}, D_{\gamma\dot\delta}\} &= i
C_{\alpha\gamma}C_{\dot\beta\dot\delta}{\partial\over\partial z}, \cr}
\eqn\maracuya $$ while the rest of the commutators do not change.

We are now in the position to define the massive multiplet. The superfield
defining the $N=2$ hypermultiplet is a scalar superfield in the two-dimensional
representation of the internal group $SU(2)_\lai$. This implies that the
defining superfield in the twisted theory has spin 1/2. We will denote the
corresponding spinor superfield by $\Phi_{\alpha}$. This superfield satisfies
certain superspace constraints which are easily derived from the constraints
satisfied in $N=2$ supersymmetry [\sohn]: $$ \big[ C_{(\alpha|\beta}D +
D_{(\alpha|\beta} \big]\Phi_{\gamma)} = 0, \,\,\,\,\,\,\,\,\,\,\,\,\,\,\,\,
D_{(\alpha\dot\beta}\Phi_{\gamma)} = 0. \eqn\chuchamel $$ Besides this
superfield we will denote by $\overline{\Phi}_{\alpha}$ the complex conjugate
spinor superfield. This superfield satisfies also the constraints \chuchamel,
$$ \big[ C_{(\alpha|\beta}D + D_{(\alpha|\beta} \big]\overline\Phi_{\gamma)} =
0, \,\,\,\,\,\,\,\,\,\,\,\,\,\,\,\,
D_{(\alpha\dot\beta}\overline\Phi_{\gamma)} = 0. \eqn\chuchameldos $$ Notice
that the theory is certainly chiral and that the choice of one chirality over
the other is dictated by the twist chosen in \tres. In other words, the
defining superfields (which before the twisting transformed as the (0,0,1/2)
representation of $SU(2)_\ele\times SU(2)_\ere\times SU(2)_\lai$) transform
under the (0,1/2) representation of $SU(2)_\ele\times SU(2)_\ere'$. The
superfields $\Phi_\alpha$ and $\overline\Phi_\alpha$ have ghost number 0.

The constraints \chuchamel\ imply that there are not component fields with spin
higher than 1/2. This fact can be easily demonstrated after working out the
following useful identities which are a consequence of the constraints
\chuchamel\ and the algebra \maracuya: $$\eqalign{ \big( C_{\alpha\beta}D +
D_{\alpha\beta} \big)\Phi_{\gamma} &= 2C_{\alpha\gamma}D\Phi_{\beta}, \cr
\big( C_{\alpha\beta}D + D_{\alpha\beta} \big)D\Phi_{\gamma} &= -i
C_{\beta\gamma}\partial\Phi_{\alpha}, \cr D_{\alpha\dot\beta}\Phi_{\gamma} &=
\half C_{\alpha\gamma}D_{\tau\dot\beta}\Phi^{\tau}, \cr
D_{\alpha\dot\gamma}D^{\beta\dot\gamma}\Phi^{\delta} &= -2i
C^{\beta\delta}\partial\Phi_{\alpha}, \cr
DD_{\eta}^{\,\,\,\,\,\dot\tau}\Phi^{\eta} &=
2i\partial_{\underline\eta}^{\,\,\,\,\,\underline{\dot\tau}}\Phi^{\eta}, \cr
D_{\alpha\gamma}D_{\eta}^{\,\,\,\,\,\dot\tau}\Phi^{\eta} &=
-2i\partial_{\underline{(\alpha}}^{\,\,\,\,\,\underline{\dot\tau}}
\Phi_{\gamma)},\cr
\partial D\Phi_{\alpha} &=
-\half\partial_{\underline{\alpha\dot\tau}}D_{\eta}^{\,\,\,\,\,\dot\tau}
\Phi^{\eta},
\cr  \partial D_{\tau\dot\beta}\Phi^{\tau} &= 4
\partial_{\underline{\tau\dot\beta}}D\Phi^{\tau}, \cr \partial^2 \Phi_{\gamma}
&= \partial_{\underline{\alpha\dot\beta}}
\partial^{\underline{\alpha\dot\beta}} \Phi_{\gamma}. \cr} \eqn\mandioca $$
One observes that all the possible higher spin components can be expressed in
terms of lower ones. Furthermore, the last equation, which can be rewritten as
the condition $P^2 + Z^2=0$, truncates the infinite expansion of the superfield
$\Phi_\alpha$ in powers of $z$, remaining a finite number of component fields.
Of course, a similar set of relations as the ones in \mandioca\ holds for
$\overline \Phi_\alpha$.

\newdimen\tableauside\tableauside=1.0ex
\newdimen\tableaurule\tableaurule=0.4pt \newdimen\tableaustep
\def\phantomhrule#1{\hbox{\vbox to0pt{\hrule height\tableaurule width#1\vss}}}
\def\phantomvrule#1{\vbox{\hbox to0pt{\vrule width\tableaurule height#1\hss}}}
\def\sqr{\vbox{%
  \phantomhrule\tableaustep
  \hbox{\phantomvrule\tableaustep\kern\tableaustep\phantomvrule\tableaustep}%
  \hbox{\vbox{\phantomhrule\tableauside}\kern-\tableaurule}}}
\def\squares#1{\hbox{\count0=#1\noindent\loop\sqr
  \advance\count0 by-1 \ifnum\count0>0\repeat}}
\def\tableau#1{\vcenter{\offinterlineskip
  \tableaustep=\tableauside\advance\tableaustep by-\tableaurule
  \kern\normallineskip\hbox
    {\kern\normallineskip\vbox
      {\gettableau#1 0 }%
     \kern\normallineskip\kern\tableaurule}%
  \kern\normallineskip\kern\tableaurule}} \def\gettableau#1
{\ifnum#1=0\let\next=\null\else
  \squares{#1}\let\next=\gettableau\fi\next}

\tableauside=2.0ex \tableaurule=0.4pt


Component fields are defined introducing adequate superspace derivatives. From
\mandioca\ follows that the only independent component fields are: $$ \eqalign{
\Phi_{\alpha} \big| &= H_{\alpha}, \cr D\Phi_{\alpha} \big| &=
2^{-1/4}u_{\alpha}, \cr D_{\rho\dot\alpha} \Phi^{\rho} \big| &= 2^{5/4}
v_{\dot\alpha}, \cr \partial\Phi_{\alpha} \big| &= i2^{1/2} K_{\alpha},\cr}
\qquad \eqalign{ \overline{\Phi}_{\alpha} \big| &= \overline{H}_{\alpha}, \cr
 D\overline{\Phi}_{\alpha} \big| &=2^{-1/4}\overline{u}_{\alpha},\cr
D_{\rho\dot\alpha} \overline{\Phi}^{\rho} \big| &=2^{5/4}
\overline{v}_{\dot\alpha}, \cr \partial\overline{\Phi}_{\alpha} \big| &=
i2^{1/2} \overline{K}_{\alpha}. \cr} \eqn\tamarindo $$ Again, the numerical
factors  in these definitions are chosen for later convenience. The
$Q$-transformations of the component fields are easily obtained using \pomelo,
\maracuya\ and \mandioca. They turn out to be: $$ \eqalign{ \delta H_{\alpha}
&= \epsilon u_{\alpha}, \cr \delta u_{\alpha} &= -\epsilon K_{\alpha}, \cr
\delta v_{\dot\alpha} &=i\epsilon\partial_{{\rho\dot\alpha}}H^{\rho},\cr
\delta K_{\alpha} &=i\epsilon \partial_{{\alpha\dot\tau}} v^{\dot\tau}. \cr}
\eqn\castana $$ In a similar way the $H_{\alpha\beta}$ and
$G_{\alpha\dot\beta}$ transformations are worked out: $$ \eqalign{ \delta'
H_{\alpha} &=2\epsilon^{\beta\gamma}C_{\beta\alpha}u_{\gamma}, \cr \delta'
u_{\alpha} &=2\epsilon^{\beta\gamma}C_{\beta\alpha}K_{\gamma}, \cr
\delta'v_{\dot\alpha}
&=-2i\epsilon^{\beta\gamma}\partial_{\beta\dot\alpha}H_{\gamma}      , \cr
\delta'K_{\alpha}&=2i\epsilon^{\beta\gamma}C_{\beta\alpha}
\partial_{\gamma\dot\beta}v^{\dot\beta}        ,\cr} \qquad \eqalign{ \delta''
H_{\alpha} &=\epsilon^{\beta\dot\gamma}C_{\beta\alpha}v_{\dot\gamma}, \cr
\delta'' u_{\alpha}
&=i\epsilon^{\beta\dot\gamma}\partial_{\alpha\dot\gamma}H_{\beta}, \cr
\delta'' v_{\dot\alpha}
&=\epsilon^{\beta\dot\gamma}C_{\dot\gamma\dot\alpha}K_{\beta}      , \cr
\delta''K_{\alpha} &=-i
\epsilon^{\beta\dot\gamma}C_{\beta\alpha}\partial_{\delta\dot\gamma}u^{\delta}
         . \cr} \eqn\castanon $$ Finally the $z$-transformations become, $$
\eqalign{ \delta_{z} H_{\alpha} &= - z K_{\alpha}, \cr \delta_{z} u_{\alpha}
&= -iz \partial_{\alpha\dot\alpha} v^{\dot\alpha},\cr  \delta_{z}
v_{\dot\alpha}&=iz \partial_{\alpha\dot\alpha}u^{\alpha},\cr \delta_{z}
K_{\alpha} &= z \tableau{1} H_\alpha. \cr} \eqn\yago $$ In this equation $z$
is a parameter and should not be confused with the coordinate $z$ introduced in
the extended superspace as in \pomelo. In these last sets of transformations
the underlines of  vector indices in partial derivatives have been removed
since after the projection there are not anticommuting vector indices. Of
course, a similar set of transformations holds for the overlined fields. The
ghost numbers of the component matter fields can be obtained easily from
\tamarindo\ and the fact that the superfield $\Phi_\alpha$ has ghost number 0.
It turns out that the set of matter fields
$(H_\alpha,u_\alpha,v_{\dot\alpha},K_\alpha)$ and $(\overline
H_\alpha,\overline u_\alpha,\overline v_{\dot\alpha},\overline K_\alpha)$ have
both ghost numbers $(0,1,-1,2)$. These ghost numbers are the charges of the
$Z_4$ symmetry of the extended topological algebra.

Since central charges are present in the theory there is not a natural measure
to construct invariant actions. Certainly, one must require $Z$-invariance and
this is not guaranteed by measures as the ones taken in \coco\ and \cocouno.
Actually, it is rather hard to find $Z$-invariant actions. Guided by the
formulation of $N=2$ supersymmetry in [\sohn] there are at least two quantities
invariant under all the symmetries of the extended topological algebra \pomelo.
These lead to the following terms entering the action for the topological
hypermultiplet,    $$ \eqalign{ {\cal L}_0^\laf & = \int d^4x \big[D^2
C_{\alpha}^{\,\,\beta} - D^{\sigma\beta}D_{\sigma\alpha}
+D_{\alpha\dot\sigma}D^{\beta\dot\sigma} \big]\big(\overline{\Phi}^{\alpha}
\buildrel\leftrightarrow\over\partial\Phi_{\beta}\big)|, \cr {\cal L}_m^\laf
&=\int d^4x \big[D^2 C_{\alpha}^{\,\,\beta} - D^{\sigma\beta}D_{\sigma\alpha}
+D_{\alpha\dot\sigma}D^{\beta\dot\sigma}
\big]\big(\overline{\Phi}^{\alpha}\Phi_{\beta}\big)|, \cr } \eqn\artocarpus $$
where the superscript ``f" stands for free. The first quantity contains the
kinetic part while the second correspond to a mass term. The action ${\cal
L}_0^\laf $ has ghost number 0 and thus models based on this action possess a
$Z_4$ symmetry. However, the action  ${\cal L}_m^\laf $ has ghost number 2.
This implies that in models where  $m\neq 0$ the ghost number symmetry is
broken to $Z_2$.

The most general form of the full action is, after writing \artocarpus\ in
terms of the component fields  \tamarindo, $$ \eqalign{ {\cal L}^\laf = &
{\cal L}^\laf_0 + m{\cal L}^\laf_m \cr = &\int d^4x
\Big[\overline{H}^{\alpha}\tableau{1}
H_{\alpha}+i\overline{u}^{\alpha}\partial_{\alpha\dot\alpha}v^{\dot\alpha}-
i\overline{v}^{\dot\alpha}\partial_{\dot\alpha\alpha}u^{\alpha}
+\overline{K}^{\alpha}K_{\alpha} \cr &
\,\,\,\,\,\,\,\,\,\,\,\,\,\,\,\,\,\,\,\,\,\,\,
+m\big(\overline{K}^{\alpha}H_{\alpha} - \overline{H}^{\alpha}K_{\alpha}\big)
+m\big(\overline{u}^{\alpha}u_{\alpha}
+\overline{v}^{\dot\alpha}v_{\dot\alpha}\big)\Big],\cr} \eqn\nuezmoscada $$
where $m$ is an arbitrary  mass parameter.

The following redefinition of the auxiliary fields $\overline K$ and $K$
isolates a mass term for $\overline H$ and $H$: $$
\eqalign{\overline{K}^{\alpha} &=\overline{K'}^{\alpha}+
m\overline{H}^{\alpha},
\cr K_{\alpha} &= {K'}_{\alpha}- m H_{\alpha}. \cr} \eqn\Recesvinto $$ The
final expression for the matter action is: $$ \eqalign{ {\cal L}^\laf = &
{\cal L}^\laf_0 + m{\cal L}^\laf_m \cr
 = &\int d^4x \Big[\overline{H}^{\alpha}\tableau{1}
H_{\alpha}+i\overline{u}^{\alpha}\partial_{\alpha\dot\alpha}v^{\dot\alpha}-
i\overline{v}^{\dot\alpha}\partial_{\dot\alpha\alpha}u^{\alpha}
+\overline{K'}^{\alpha}{K'}_{\alpha} \cr &
\,\,\,\,\,\,\,\,\,\,\,\,\,\,\,\,\,\,\,\,\,\,+m^2
\overline{H}^{\alpha}H_{\alpha}+ m\big(\overline{u}^{\alpha}u_{\alpha}
+\overline{v}^{\dot\alpha}v_{\dot\alpha}\big)\Big].\cr}  \eqn\Chindasvinto $$

So far we have formulated the theory on a flat four-dimensional manifold. In
order to construct the topological model we must rewrite the theory for an
arbitrary curved space, in other words, we must introduce a background metric
$g_{\mu\nu}$ on the manifold and covariantize the formulation. Furthermore,
since there are spinorial fields, the manifold chosen must be a spin manifold.
We will assume that some choice of spin structure has been made. This process
presents some surprises since, as we now describe, one is forced to introduce
new terms in the action which depend on the curvature associated to
$g_{\mu\nu}$ in order to maintain the symmetries of the theory. Looking back to
the derivation of the relations \mandioca, it turns our that for an arbitrary
curved space the last relation becomes, $$ \partial^2\Phi_{\gamma} =
2\big(\tilde{\tableau{1}} +{1\over4}R \big)\Phi_{\gamma}, \eqn\axioma $$ where
$\tilde{\tableau{1}}$ is the covariant laplacian and $R$ is the scalar
curvature. This has important consecuences in the new form of the
$z$-transformations \yago\ and in the new form of the action \Chindasvinto. On
the other hand, the rest of the transformations, namely, the ones in \castana\
and \castanon\ remain the same once the partial derivatives are replaced by
covariant ones. The $z$-transformations \yago\ become, $$ \eqalign{ \delta_{z}
H_{\alpha} &= -z K_{\alpha}, \cr \delta_{z} u_{\alpha} &=
-iz\tilde\nabla_{\alpha\dot\alpha} v^{\dot\alpha},\cr \delta_{z}
v_{\dot\alpha} &=iz \tilde\nabla_{\alpha\dot\alpha}u^{\alpha},\cr  \delta_{z}
K_{\alpha} &= z (\tilde{\tableau{1}}+{1\over4}R) H_\alpha, \cr} \eqn\yagodos
$$ where $\tilde\nabla_{\alpha\dot\alpha}$ denotes the covariant derivative.
The final form of the action is, $$ \eqalign{ {\cal L}^\lag= & {\cal L}^\lag_0
+ m{\cal L}^\lag_m \cr
  = &\int_{M} d^4x \sqrt{g} \Big[
\overline{H}^{\alpha}(\tilde{\tableau{1}}+{1\over4}R) H_{\alpha}
 +i\overline{u}^{\alpha}\tilde\nabla_{{\alpha\dot\alpha}}{v}^{\dot\alpha} -
{i}\overline{v}^{\dot\alpha}\tilde\nabla_{{\alpha\dot\alpha}}u^{\alpha}
+\overline{K'}^{\alpha}K_{\alpha}'\cr
&\,\,\,\,\,\,\,\,\,\,\,\,\,\,\,\,\,\,\,\,\,\,\,\,\,\,\,\,\,\, + m^2
\overline{H}^\alpha H_\alpha +m(\overline{u}^\alpha u_\alpha +
\overline{v}^{\dot\alpha}v_{\dot\alpha})\Big].\cr} \eqn\maria $$ The
superscript g in \maria\ indicates that a choice of metric and spin structure
has been made. This action is invariant under $Q$ and $z$ transformations.
However, it is not invariant under $H_{\alpha\beta}$ and $G_{\alpha\dot\beta}$
transformations unless their corresponding parameters satisfy, $$
\tilde\nabla_{\alpha\dot\beta} \epsilon_{\gamma\sigma} =0,
\,\,\,\,\,\,\,\,\,\,\,\,\,\,\, \tilde\nabla_{\alpha\dot\beta}
\epsilon_{\gamma\dot\sigma} =0. \eqn\turroncete $$ Certainly, not all spin
manifolds admit covariantly constant vectors or self-dual tensors. Therefore,
those symmetries do not exist in general. However, as discussed before, these
symmetries might be useful to construct the coupling to topological gravity.

Using the covariantized version of \castana\ it is simple to verify that  for
$m=0$ the action \maria\ is $Q$-exact. Notice that contrary to the case of
topological Yang-Mills, this is not obvious from the form of the superspace
action \artocarpus. In fact, it turns out that, $$ {\cal L}^\lag_0 = \big\{ Q,
\tilde\Lambda^\lag  \big\}, \eqn\adri $$ where, $$ \tilde\Lambda^\lag =\half
\int_{M} d^4x \sqrt{g} \Big[ i\overline{H}^\alpha
\tilde\nabla_{\alpha\dot\alpha} v^{\dot\alpha}
+i\overline{v}^{\dot\alpha}\tilde\nabla_{\alpha\dot\alpha}H^{\alpha}
-\overline{K}^\alpha u_\alpha - \overline{u}^{\alpha}K_{\alpha}\Big].
\eqn\adriana $$ The  invariance under $Q$ and $Z$ of ${\cal L}^\lag_0$ can be
regarded simply as a consequence of \adri\ and the relations: $$ [Q,Z]=0,
\,\,\,\,\,\,\,\,\,\,\,\,\,\,\,\,\,\,\,\,\, [Z,\tilde\Lambda^\lag]=0. \eqn\laura
$$

Equation \adri\ does not ensure that the model we have built is topological,
since from the $Q$-exactness of the whole action it does not  follow that the
energy-momentum tensor is also $Q$-exact. This would be so if the
$Q$-transformations would not contain covariant derivatives, which is not our
case. Nevertheless we show below that the non-$Q$-exact part of the
energy-momentum tensor vanishes on-shell. On the other hand, the $Q$-exactness
of the action makes the theory exact in the small coupling limit. These two
facts suffice to render the theory topological. This implies [\tqft] that the
vacuum expectation values of operators which are invariant under
$Q$-transformations lead to topological invariant quantities. A description of
the observables of this theory was presented in [\bts]. These turn out to be a
very restrictive set because of the strong conditions imposed by
$z$-invariance. As we will observe in the discussion concerning the coupling of
these models to topological Yang-Mills in the next section, the $z$-symmetry
is so restrictive that
the only observables are the ones of the form \observables.

\endpage
\chapter{Matter coupling to Donaldson-Witten theory}

We will construct the coupling of the topological hypermultiplet of the
previous section to Donaldson-Witten theory covariantizing the algebra of the
extended superspace derivatives, imposing the gauge constraints \catorce, and
covariantizing the defining constraints on the matter superfields. Let us
therefore pick a gauge group $G$ and introduce gauge connections as in \fresa.
The form of the constraints \catorce\ now becomes: $$ \eqalign{ \{\D, \D \}
&=i\big(\partial-\half V\big), \cr \{\D_{\alpha\beta}, \D_{\gamma\delta} \}
&=iC_{(\alpha| \gamma}C_{\beta)\delta}\big(\partial-\half V\big), \cr
\{\D_{\alpha\dot\beta}, \D_{\gamma\dot\delta} \}
&=iC_{\alpha\gamma}C_{\dot\beta\dot\delta}\big(\partial-W\big), \cr}
\eqn\bellota $$ where the scalar superfields $V$ and $W$ are the same as in
\catorce. The remaining fundamental (anti)commutation relations do not change.
As the central charge commutes with everything, the gauge sector of the theory
is as before (the Bianchi identities remain unchanged). On the other hand, the
matter sector has to be reconsidered.

Let us consider a commuting spinor superfield $\Phi_\alpha$ which transforms
under a representation of the gauge group $G$, and another spinor superfield
$\overline\Phi^\alpha$ which transform under the conjugate representation. The
covariant form of the constraints \chuchamel, which are the defining equations
of the hypermultiplet, now read, $$ \big[C_{(\alpha|\beta}\D +
\D_{(\alpha|\beta} \big]\Phi_{\gamma)} = 0, \,\,\,\,\,\,\,\,\,\,\,\,\,\,\,\,
\D_{(\alpha\dot\beta}\Phi_{\gamma)} = 0. \eqn\pexego $$ The equivalent of the
identities \mandioca, which determine the set of independent component fields,
consist of the set of equations: $$ \eqalign{ \big( C_{\alpha\beta}\D +
\D_{\alpha\beta} \big)\Phi_{\gamma} =& 2C_{\alpha\gamma}\D\Phi_{\beta}, \cr
\D_{\alpha\dot\beta}\Phi_{\gamma} =& \half
C_{\alpha\gamma}\D_{\tau\dot\beta}\Phi^{\tau}, \cr
\D_{\alpha\dot\gamma}\D_{\delta}^{\,\,\dot\gamma}\Phi^{\delta} =&
4i\big(W-\partial\big)\Phi_{\alpha}, \cr \D\D_{\eta}^{\,\,\dot\tau}\Phi^{\eta}
=& 2i\D_{\underline\eta}^{\,\,\,\,\,\underline{\dot\tau}}\Phi^{\eta}, \cr
\D_{\alpha\gamma}\D_{\eta}^{\dot\tau}\Phi^{\eta} =&
-2i\D_{\underline{(\alpha}}^{\,\,\,\,\,\underline{\dot\tau}}\Phi_{\gamma)}, \cr
\partial\D\Phi_{\alpha}
=&-\half\D_{\underline{\alpha\dot\tau}}\D_{\eta}^{\,\,\dot\tau}\Phi^{\eta}
+W\D\Phi_{\alpha} +\half \D W\Phi_{\alpha} +\half\D_{\alpha\tau}W\Phi^{\tau},
\cr \partial\D_{\tau\dot\beta}\Phi^{\tau} =&
4\D_{\underline{\tau\dot\beta}}\D\Phi^{\tau} +\half
V\D_{\tau\dot\beta}\Phi^{\tau} +\D_{\tau\dot\beta}V\Phi^{\tau}, \cr \partial^2
\Phi_{\alpha} =& 2\tableau{1}\Phi_{\alpha}
+{i\over4}\D_{\alpha\dot\beta}V\D_{\tau}^{\,\,\dot\beta}\Phi^{\tau} -i\D W
\D\Phi_{\alpha}, \cr &+i\D_{\alpha\tau}W\D\Phi^{\tau} + W\partial\Phi_{\alpha}
+\half V\partial\Phi_{\alpha}\cr & +iG_{\alpha\tau}\Phi^{\tau}
-{1\over4}\big{\{} V,W \big{\}}\Phi_{\alpha}. \cr} \eqn\gominola $$ We will
carry out a covariant projection into component fields. This means that
component fields must be defined as in \tamarindo\ replacing ordinary
superspace derivatives by covariant ones: $$ \eqalign{ \Phi_{\alpha} \big|
&=H_{\alpha}, \cr \D\Phi_{\alpha} \big| &=2^{-1/4}u_{\alpha}, \cr
\D_{\rho\dot\alpha} \Phi^{\rho} \big| &= 2^{5/4} v_{\dot\alpha}, \cr
\partial\Phi_{\alpha} \big| &=i2^{1/2} K_{\alpha},\cr} \qquad \eqalign{
\overline{\Phi}_{\alpha} \big| &=\overline{H}_{\alpha}, \cr
\D\overline{\Phi}_{\alpha} \big| &=2^{-1/4}\overline{u}_{\alpha},\cr
\D_{\rho\dot\alpha} \overline{\Phi}^{\rho} \big|
&=2^{5/4}\overline{v}_{\dot\alpha}, \cr \partial\overline{\Phi}_{\alpha} \big|
&= i2^{1/2}\overline{K}_{\alpha}. \cr} \eqn\tamarindodos $$ The transformation
laws of these component fields are derived making use of \sandia\ and the
relations \gominola. For $Q$-transformations these are: $$ \eqalign{ \delta
H_{\alpha} =& \epsilon u_{\alpha}, \cr \delta u_{\alpha} =&
-\epsilon\big(K_{\alpha} + i\phi H_{\alpha}\big), \cr \delta v_{\dot\alpha} =&
i\epsilon\D_{{\alpha\dot\alpha}} H^{\alpha}, \cr \delta K_{\alpha} =&
i\epsilon\big(\D_{\alpha\dot\alpha} v^{\dot\alpha} -\lambda u_{\alpha},\cr &
-\eta H_{\alpha} -\chi_{\alpha\beta} H^{\beta}\big), \cr} \qquad \eqalign{
\delta \overline{H}_{\alpha} =& \epsilon \overline{u}_{\alpha}, \cr \delta
\overline{u}_{\alpha} =& -\epsilon\big(\overline{K}_{\alpha}
-i\phi\overline{H}_{\alpha}\big), \cr \delta \overline{v}_{\dot\alpha} =&
i\epsilon\D_{{\alpha\dot\alpha}}\overline{H}^{\alpha}, \cr
\delta\overline{K}_{\alpha} =&
i\epsilon\big(\D_{\alpha\dot\alpha}\overline{v}^{\dot\alpha} +\lambda
\overline{u}_{\alpha},\cr & +\eta\overline{H}_{\alpha} +\chi_{\alpha\beta}
\overline{H}^{\beta}\big), \cr} \eqn\Jeremiah  $$ while for
$H_{\alpha\beta}$-transformations, $$ \eqalign{ \delta' H_{\alpha}
&=2\epsilon^{\gamma\delta}C_{\gamma\alpha}u_{\delta}, \cr \delta' u_{\alpha}
&=2\epsilon^{\gamma\delta}C_{\gamma\alpha}\big(K_{\delta}+i\phi
H_{\delta}\big), \cr \delta'v_{\dot\alpha}
&=-2i\epsilon^{\gamma\delta}\D_{\gamma\dot\alpha}H_{\delta}      , \cr
\delta'K_{\alpha}&=2i\epsilon^{\gamma\delta}C_{\gamma\alpha}\big(
\D_{\delta\dot\beta}v^{\dot\beta}-\lambda u_{\delta}\cr
&\,\,\,\,\,\,\,\,\,\,-\eta H_{\delta}-\chi_{\delta\beta}H^{\beta}\big)
,\cr}  \qquad  \eqalign{ \delta' \overline{H}_{\alpha}
&=2\epsilon^{\gamma\delta}C_{\gamma\alpha}\overline{u}_{\delta}     , \cr
\delta'\overline{u}_{\alpha}
&=2\epsilon^{\gamma\delta}C_{\gamma\alpha}\big(K_{\delta}-i\phi
H_{\delta}\big)     , \cr \delta' \overline{v}_{\dot\alpha}
&=-2i\epsilon^{\gamma\delta}\D_{\gamma\dot\alpha}\overline{H}_{\delta} , \cr
\delta'\overline{K}_{\alpha} &=2i\epsilon^{\gamma\delta}C_{\gamma\alpha}\big(
\D_{\delta\dot\beta}\overline{v}^{\dot\beta}+\lambda u_{\delta}\cr &
\,\,\,\,\,\,\,\,\,\,+\eta\overline{
H}_{\delta}+\chi_{\delta\beta}\overline{H}^{\beta}\big)      , \cr}
\eqn\melonazo  $$ and, finally, for $G_{\alpha\dot\beta}$ transformations, $$
\eqalign{ \delta'' H_{\alpha}
&=\epsilon^{\gamma\dot\delta}C_{\gamma\alpha}v_{\dot\delta}, \cr \delta''
u_{\alpha} &=i\epsilon^{\gamma\dot\delta}\D_{\alpha\dot\delta}H_{\gamma} ,\cr
\delta'' v_{\dot\alpha}
&=\epsilon^{\gamma\dot\delta}C_{\dot\delta\dot\alpha}\big(K_{\gamma}+i\lambda
H_{\gamma}\big)      , \cr  \delta'' K_{\alpha} &=-i
\epsilon^{\gamma\dot\delta}C_{\gamma\alpha}\big(\D_{\rho\dot\delta}u^{\rho}\cr
& \,\,\,\,\,\,\,\,\,\,-i\psi_{\rho\dot\delta}H^{\rho}+\phi
v_{\dot\delta}\big)          , \cr} \qquad  \eqalign{  \delta''
\overline{H}_{\alpha}
&=\epsilon^{\gamma\dot\delta}C_{\gamma\alpha}\overline{v}_{\dot\delta}     ,
\cr \delta''\overline{u}_{\alpha}
&=i\epsilon^{\gamma\dot\delta}\D_{\alpha\dot\delta} \overline{H}_{\gamma}
, \cr  \delta''\overline{v}_{\dot\alpha}
&=\epsilon^{\gamma\dot\delta}C_{\dot\delta\dot\alpha}\big(K_{\gamma}-i\lambda
H_{\gamma}\big)  , \cr  \delta''\overline{K}_{\alpha} &=-i
\epsilon^{\gamma\dot\delta}C_{\gamma\alpha}\big(\D_{\rho\dot\delta}
\overline{u}^{\rho}\cr
&\,\,\,\,\,\,\,\,\,\,+i\psi_{\rho\dot\delta}H^{\rho}-\phi
v_{\dot\delta}\big)      . \cr}   \eqn\melonaza  $$ On the other hand, it is
also important to work out the form of the $z$-transformations, these become:
$$ \eqalign{ \delta_z {H}_{\alpha} =&  - z K_{\alpha}, \cr \delta_z
{u}_{\alpha} =&  -iz \big(\D_{\alpha\dot\alpha}v^{\dot\alpha}- \lambda
u_{\alpha}-\eta H_{\alpha}-\chi_{\alpha\beta}H^{\beta}\big),\cr \delta_z
{v}_{\dot\alpha} =&iz\big(\D_{\alpha\dot\alpha}u^{\alpha}+\phi v_{\dot\alpha}-
i\psi_{\alpha\dot\alpha}H^{\alpha}\big),\cr \delta_z {K}_{\alpha} =&
z\Big(\tableau{1}H_\alpha
+\psi_{\alpha\dot\alpha}v^{\dot\alpha}-\half\big\{\lambda, \phi \big\} -i\eta
u_{\alpha}+i\chi_{\alpha\beta}H^{\beta} \cr
&+{i\over2}G_{\alpha\beta}H^{\beta}+i\lambda K_{\alpha}+i\phi
K_{\alpha}\Big),\cr} \eqn\trillo $$ $$ \eqalign{ \delta_z
\overline{H}_{\alpha} =&  -z\overline K_{\alpha} , \cr \delta_z \overline
{u}_{\alpha} =& -iz \big(\D_{\alpha\dot\alpha}\overline v^{\dot\alpha}+\lambda
\overline u_{\alpha}+\eta \overline H_{\alpha}+\chi_{\alpha\beta}\overline
H^{\beta}\big),\cr \delta_z \overline {v}_{\dot\alpha} =&
iz\big(\D_{\alpha\dot\alpha}\overline u^{\alpha}-\phi
\overline{v}_{\dot\alpha} +i\psi_{\alpha\dot\alpha}\overline
H^{\alpha}\big),\cr \delta_z \overline {K}_{\alpha} =& z\Big(
\tableau{1}\overline H_\alpha -\psi_{\alpha\dot\alpha}\overline
v^{\dot\alpha}-\half\big\{\lambda, \phi \big\} +i\eta \overline
u_{\alpha}-i\chi_{\alpha\beta}H^{\beta}\cr
&-{i\over2}G_{\alpha\beta}H^{\beta}-i\lambda K_{\alpha}-i\phi K_{\alpha}
\Big).\cr} \eqn\trillodos $$ In the transformations \Jeremiah, \melonazo,
\melonaza, \trillo\ and \trillodos\ commuting vector indices have not been
underlined since at the component level there is not risk to be mistaken.

The coupled matter action turns out to be made out of the covariantization of
the  terms in \artocarpus: $$ \eqalign{ {\cal L}^\la_0 &= \int d^4x \big[\D^2
C_{\alpha}^{\,\,\beta} - \D^{\sigma\beta}\D_{\sigma\alpha}
+\D_{\alpha\dot\sigma}\D^{\beta\dot\sigma} \big]\big(\overline{\Phi}^{\alpha}
\buildrel\leftrightarrow\over\partial\Phi_{\beta}\big)|, \cr {\cal L}^\la_m &=
\int d^4x \big[\D^2 C_{\alpha}^{\,\,\beta} - \D^{\sigma\beta}\D_{\sigma\alpha}
+\D_{\alpha\dot\sigma}\D^{\beta\dot\sigma}
\big]\big(\overline{\Phi}^{\alpha}\Phi_{\beta}\big)|, \cr}
 \eqn\Isaiah $$ where the superscript tYM indicates that the fields of
topological Yang-Mills have been considered in the action \Isaiah\ as
background fields. Before writing the action in components let us analyze the
form of the symmetry transformations when the theory is placed on an arbitrary
spin manifold endowed with a metric $g_{\mu\nu}$. All transformations in
\Jeremiah, \melonazo\ and \melonaza\ remain the same after the replacement of
the Yang-Mills covariant derivative $\nabla_{\alpha\dot\beta}$ by the full
covariant derivative ${\cal D}_{\alpha\dot\beta}$ introduced in \turron. On
the other hand, since the last equation of \gominola\ gets a term involving
the scalar curvature, $$ \eqalign{ \partial^2 \Phi_{\alpha} =&
2\big(\tableau{1}+{1\over4}R\big)\Phi_{\alpha} +{i\over4}{\cal
D}_{\alpha\dot\beta}V{\cal D}_{\tau}^{\,\,\dot\beta}\Phi^{\tau} -i\D W
\D\Phi_{\alpha}, \cr &+i{\cal D}_{\alpha\tau}W\D\Phi^{\tau} +
W\partial\Phi_{\alpha} +\half V\partial\Phi_{\alpha}\cr &
+iG_{\alpha\tau}\Phi^{\tau} -{1\over4}\big{\{} V,W \big{\}}\Phi_{\alpha}, \cr}
\eqn\bolzano $$ the $z$-transformation of $K_\alpha$ and $K_\alpha'$ become
modified in the following form: $$ \eqalign{ \delta_z {K}_{\alpha} =&
z\Big[\big(\tableau{1}+{1\over4}R\big)H_\alpha
+\psi_{\alpha\dot\alpha}v^{\dot\alpha}-\half\big\{\lambda, \phi \big\} -i\eta
u_{\alpha}+i\chi_{\alpha\beta}H^{\beta} \cr
&+{i\over2}G_{\alpha\beta}H^{\beta}+i\lambda K_{\alpha}+i\phi K_{\alpha}\Big],
\cr \delta_z \overline {K}_{\alpha} =& z\Big[
\big(\tableau{1}+{1\over4}R\big)\overline H_\alpha
-\psi_{\alpha\dot\alpha}\overline v^{\dot\alpha}-\half\big\{\lambda, \phi
\big\} +i\eta \overline u_{\alpha}-i\chi_{\alpha\beta}H^{\beta}\cr
&-{i\over2}G_{\alpha\beta}H^{\beta}-i\lambda K_{\alpha}-i\phi K_{\alpha}
\Big].\cr} \eqn\heine $$ In \bolzano\ and \heine\ $\tableau{1}$ represents the
full covariant laplacian. The rest of the $z$-transformations in \trillo\ and
\trillodos\  have the same form once the replacement
$\nabla_{\alpha\dot\beta}\rightarrow {\cal D}_{\alpha\dot\beta}$ is performed.

{}From these terms the full action is defined as in \nuezmoscada. Writing it in
terms of the component fields and redefining the auxiliary fields $K$ and
$\overline{K}$ as in \Recesvinto\ one finds, $$ \eqalign{ {\cal L}^\laga = &
{\cal L}^\laga_0  + m {\cal L}^\laga_m \cr = & \int_M d^4x \sqrt g
\Big[\overline{K'}^{\alpha}{K'}_{\alpha}
+\overline{H}^{\alpha}\big(\tableau{1} +{1\over4}R\big)H_{\alpha}
+{i\over2}\overline{H}^{\alpha}F_{\alpha\beta}^{+}H^{\beta}+
i\overline{u}^{\alpha}{\cal
D}_{\alpha\dot\alpha}v^{\dot\alpha}\cr &- i\overline{v}^{\dot\alpha}{\cal
D}_{\dot\alpha\alpha}u^{\alpha}
+\overline{H}^{\alpha}\psi_{\alpha\dot\alpha}v^{\dot\alpha}
-\overline{v}^{\dot\alpha}\psi_{\dot\alpha\alpha}H^{\alpha}
-i\overline{H}^{\alpha}\eta u_{\alpha} -i\overline{u}^{\alpha}\eta H_{\alpha}
+i\overline{H}^{\alpha}\chi_{\alpha\beta}u^{\beta}\cr &
-i\overline{u}^{\alpha}\chi_{\alpha\beta}H^{\beta}
-i\overline{u}^{\alpha}\lambda u_{\alpha} -i\overline{v}^{\dot\alpha}\phi
v_{\dot\alpha} +{i\over2}\overline{H}^{\alpha}G_{\alpha\beta}H^{\beta}
-\half\overline{H}^{\alpha}\big{\{}\phi, \lambda \big{\}}H_{\alpha}  \cr &
+m^2\overline{H}^{\alpha}H_{\alpha}+m\big(\overline{u}^{\alpha}u_{\alpha}
+\overline{v}^{\dot\alpha}v_{\dot\alpha}\big)
-im\overline{H}^{\alpha}\big(\phi+\lambda \big) H_{\alpha} \Big],\cr}
\eqn\Daniel $$ where $M$ denotes the four-dimensional spin manifold where the
theory is defined. This action is invariant under the full extended
topological algebra provided the parameters satisfy the relations \turron.
Certainly, in general, only the $Q$-symmetry will hold.  At this moment one
should ask if the action ${\cal L}^\laga_0$ in \Daniel\ is $Q$-exact as was
the case for the action with no coupling to topological Yang-Mills. Contrary
to the action \nuezmoscada, ${\cal L}^\laga_0$ is not $Q$-exact. In addition
it turns out that the energy-momentum tensor is not $Q$-exact and therefore it
is not clear if the  theory is topological. Examples of topological theories
whose action is not $Q$-exact but its energy momentum is are known
[\vafa,\pablo]. However, this is not the case here.

Certainly, the mass terms  of $m {\cal L}^\laga_m $ in \Daniel\ break the
topological symmetry. What is in some sense unexpected is that the action
${\cal L}^\laga_0$ also might lead to a breaking of the topological symmetry.
The analysis of this phenomena is carried out in the sect. 7. To finish this
section let us finally write the full action $S^\lag$ of the topological model
under consideration. This action is, $$ S^\lag = {\cal L}_\ym^\lag + {\cal
L}^\laga_0, \eqn\borel $$ where ${\cal L}^\laga_0$ is the action given in
\Daniel\ and ${\cal L}_\ym^\lag$ is the covariantized form of the action
\veintidos: $$ \eqalign{ {\cal L}_\ym^\lag = & {1\over e^2} \int_M d^4x
\sqrt{g} \tr \Big\{{1\over8}\big(F_{+}^2 -G^2\big)  -\chi^{\alpha\gamma}{\cal
D}_{{\alpha\dot\beta}}\psi_{\gamma}^{\,\,\dot\beta} -2\lambda\tableau{1}\phi+
2\eta{\cal D}_{{\alpha\dot\tau}}\psi^{\alpha\dot\tau} \cr
&\,\,\,\,\,\,\,\,\,\,\,\,\,\,\,\,\,\,\,\,
-i\lambda\big\{\psi_{\alpha\dot\tau},\psi^{\alpha\dot\tau}\big\}
-{i\over2}\phi\big\{\chi_{\alpha\gamma},\chi^{\alpha\gamma}\big\}
+2i\phi\big\{\eta,\eta\big\}+{\big[\lambda,\phi\big]}^2  \Big\}. \cr}
\eqn\cauchy $$ In this action the gauge field can be taken in any
representation of the gauge group, for example, one could think just in the
representation chosen for the matter fields in \Daniel. The difference
between choosing one representation or another is just a global factor which
can be reabsorbed in the coupling constant $e$. Because of the $Q$-exactness
of \cauchy\ (see \veintidos) the observables of the theory are independent of
$e$.

To show that the action ${\cal L}^\laga_0$ in \Daniel\ is not $Q$-exact one
just has to write all possible terms quadratic in matter fields with ghost
number -1. It turns out that no combination of those terms leads to ${\cal
L}^\laga_0$. The $z$-symmetry present in ${\cal L}^\laga_0$ is very
restrictive. If ${\cal L}^\laga_0$ were $Q$ of some quantity, presumably such a
quantity should be invariant under $z$-transformations. However, it does not
exist a $z$-invariant of ghost number -1 and quadratic in the matter fields. On
the other hand, it is clear from the form of the action for the free case in
\adri\ that part of the ${\cal L}^\laga_0$ is $Q$ exact. Indeed, one finds that
$$ {\cal L}^\laga_0 = \{Q,\Lambda^\lag \} + L^\laga_0, \eqn\rama $$ where, $$
\Lambda^\lag =\half \int_{M} d^4x \sqrt{g} \Big[ i\overline{H}^\alpha {\cal
D}_{\alpha\dot\alpha} v^{\dot\alpha} +i\overline{v}^{\dot\alpha}{\cal
D}_{\alpha\dot\alpha}H^{\alpha} -\overline{K}^\alpha u_\alpha
-\overline{u}^{\alpha}K_{\alpha}\Big], \eqn\nujan $$ and, $$ \eqalign{
 L^\laga_0 = \half\int_M d^4x \sqrt g \Big[&
\overline{H}^{\alpha}\psi_{\alpha\dot\alpha}v^{\dot\alpha}
-\overline{v}^{\dot\alpha}\psi_{\dot\alpha\alpha}H^{\alpha}
-i\overline{H}^{\alpha}\eta u_{\alpha} -i\overline{u}^{\alpha}\eta H_{\alpha}
\cr & +i\overline{H}^{\alpha}\chi_{\alpha\beta}u^{\beta}
-i\overline{u}^{\alpha}\chi_{\alpha\beta}H^{\beta}
-2i\overline{v}^{\dot\alpha}\phi v_{\dot\alpha}
-\overline{H}^{\alpha}\big{\{}\phi,\lambda\big{\}}H_{\alpha} \cr &
+i\overline{H}^{\alpha}G_{\alpha\beta}H^{\beta} -i\overline{K'}^{\alpha}\phi
H_{\alpha} +i \overline{H}^{\alpha}\phi K'_{\alpha} \Big].\cr} \eqn\metha $$
This part of the action seems as complicated as the original action \Daniel.
Notice, however, that in \metha\ there are only interaction vertices. The form
of the energy-momentum tensor of this theory will be studied in sect. 7.

\endpage
\chapter{The truncated theory}

So far, the models we have presented possess all the symmetries of the
topological algebra, provided that \turron\ holds. It turns out that the
resulting theory can be truncated making it simpler. This truncation consist of
disregarding the fields $\lambda$, $\eta$ and $\chi$, and the auxiliary field
$G$ in both the coupling of matter to Donaldson-Witten theory and the matter
fields transformation laws. In other words, matter is coupled to a minimal set
of gauge fields, $A_{\mu}$, its $Q$-partner $\psi_{\mu}$ and the field $\phi$.
We give now the truncated $\delta$ and $\delta ''$ transformations for this
minimal set of fields: $$ \eqalign{ \delta A_{\alpha\dot\alpha}&=\epsilon
\psi_{\alpha\dot\alpha}, \cr \delta \psi_{\alpha\dot\beta} &=-\epsilon
\cald_{{\alpha\dot\beta}} \phi, \cr \delta \phi &= 0, \cr} \qquad \eqalign{
\delta'' A_{\alpha\dot\alpha}&=0, \cr
\delta''\psi_{\alpha\dot\alpha}&={i\over2}\epsilon^{\gamma\dot\delta}
C_{\gamma\alpha}F^{-}_{\dot\alpha\dot\delta}, \cr \delta''\phi
&=-i\epsilon^{\beta\dot\gamma}\psi_{\beta\dot\gamma}. \cr} \eqn\Moses $$

It is simple to verify that the $Q$-transformations close up to gauge
transformations generated by $\phi$: $$ \eqalign{ \big[\delta_2, \delta_1\big]
A_{\alpha\dot\alpha}&= -2i\epsilon_1 \epsilon_2 \cald_{\alpha\dot\alpha}\phi
\cr \big[\delta_2, \delta_1\big] \psi_{\alpha\dot\alpha}&= 2i\epsilon_1
\epsilon_2 \big[\psi_{\alpha\dot\alpha},\phi\big] .\cr} \eqn\Aaron $$ As for
the $G$-transformations, one would expect that the commutator of two of them
would give a gauge transformation generated by $\lambda$. This is indeed what
happens, although in the truncated case this is zero since $\lambda$ has been
set to zero,
$$ \big[{\delta''}_2, {\delta''}_1\big]
A_{\alpha\dot\alpha}=\big[{\delta''}_2,  {\delta''}_1\big]
\psi_{\alpha\dot\alpha}=0. \eqn\David $$ Now we should calculate the commutator
of a $Q$ and a $G$-transformation to make sure of having a topological
algebra. It should give a derivative of the field on which the transformations
act. This comes out to be true. $$  \eqalign{
\big[\delta,\delta''\big]A_{\alpha\dot\alpha}&=-{i\over2}\epsilon\epsilon^
{\gamma\dot\delta}C_{\gamma\alpha}F^{-}_{\dot\alpha\dot\delta},\cr
\big[\delta,\delta''\big]\psi_{\alpha\dot\alpha}&=-{i\over2}\epsilon\epsilon^
{\gamma\dot\delta}C_{\gamma\alpha}\cald_{\rho(\dot\alpha}\psi_{\dot\delta)}
^{\,\,\rho},\cr \big[\delta,\delta''\big]\phi &=-i\epsilon\epsilon^
{\gamma\dot\delta}\cald_{\gamma\dot\delta}\phi .\cr} \eqn\Salomon $$ We now
give the action of the truncated $Q$ and $G$-transformations on matter fields
and explore its consistency with the topological algebra. $$ \eqalign{ \delta
H_{\alpha} =& \epsilon u_{\alpha}, \cr \delta u_{\alpha} =&
-\epsilon\big(K_{\alpha} + i\phi H_{\alpha}\big), \cr \delta v_{\dot\alpha} =&
i\epsilon\cald_{\alpha\dot\alpha} H^{\alpha}, \cr \delta K_{\alpha} =&
i\epsilon\cald_{\alpha\dot\alpha} v^{\dot\alpha}, \cr} \qquad \eqalign{
\delta'' H_{\alpha}
&=\epsilon^{\gamma\dot\delta}C_{\gamma\alpha}v_{\dot\delta}, \cr \delta''
u_{\alpha} &=i\epsilon^{\gamma\dot\delta}\cald_{\alpha\dot\delta}H_{\gamma}
,\cr \delta'' v_{\dot\alpha}
&=\epsilon^{\gamma\dot\delta}C_{\dot\delta\dot\alpha}K_{\gamma}      , \cr
\delta'' K_{\alpha} &=-i
\epsilon^{\gamma\dot\delta}C_{\gamma\alpha}
\big(\cald_{\rho\dot\delta}u^{\rho}\cr
& \,\,\,\,\,\,\,\,\,\,-i\psi_{\rho\dot\delta}H^{\rho}+\phi
v_{\dot\delta}\big).\cr} \eqn\Jephtah
$$
The commutator of two
$Q$-transformations is now a combined $Z$ and gauge transformation generated
by $\phi$:  $$ \eqalign{ \big[\delta_2, \delta_1\big] H_{\alpha}&=
-2\epsilon_1\epsilon_2\big(K_{\alpha}+i\phi H_{\alpha}\big) \cr \big[\delta_2,
\delta_1\big] u_{\alpha}&=
-2i\epsilon_1\epsilon_2\big(\cald_{\alpha\dot\alpha}v^{\dot\alpha} +\phi
u_{\alpha}\big)\cr \big[\delta_2, \delta_1\big]
v_{\dot\alpha}&=2\epsilon_1\epsilon_2\big(i\cald_{\alpha\dot\alpha}u^{\alpha}
+\psi_{\alpha\dot\alpha}H^{\alpha}\big)\cr \big[\delta_2, \delta_1\big]
K_{\alpha}&=2\epsilon_1\epsilon_2\Big[\big(\tableau{1}+{1\over4}R\big)
H_{\alpha}+{i\over2}F^{+}_{\alpha\beta}H^{\beta}+
\psi_{\alpha\dot\alpha}v^{\dot\alpha}\big). \cr} \eqn\Abraham $$ The
commutator of two $G$-transformations gives a $Z$-transformation, in analogy
with the case of Donaldson-Witten fields.  $$ \eqalign{ \big[{\delta''}_2,
{\delta''}_1\big]
H_{\alpha}&=-(\epsilon_1)^{\beta\dot\gamma}(\epsilon_2)_{\beta\dot\gamma}
K_{\alpha} \cr \big[{\delta''}_2, {\delta''}_1\big]
u_{\alpha}&=-i(\epsilon_1)^{\beta\dot\gamma}(\epsilon_2)_{\beta\dot\gamma}
\cald_{\alpha\dot\alpha}v^{\dot\alpha} \cr \big[{\delta''}_2, {\delta''}_1\big]
v_{\dot\alpha}&=i(\epsilon_1)^{\beta\dot\gamma}(\epsilon_2)_{\beta\dot\gamma}
\big(\cald_{\delta\dot\alpha}u^{\delta}+\phi v_{\dot\alpha}
-i\psi_{\delta\dot\alpha}H^{\delta}\big) \cr \big[{\delta''}_2,
{\delta''}_1\big]
K_{\alpha}&=(\epsilon_1)^{\beta\dot\gamma}(\epsilon_2)_{\beta\dot\gamma}
\Big[\big(\tableau{1}+{1\over4}R\big)H_{\alpha}+{i\over2}F^{+}_{\alpha\beta}
H^{\beta}+i\phi K_{\alpha} -\psi_{\alpha\dot\alpha}v^{\dot\alpha}\Big] .\cr}
\eqn\Ismael $$ If we now work out the commutator of a $Q$ and a
$G$-transformation, we find the following pattern: $$ \big[\delta,
\delta''\big]\Phi =
-i\epsilon\epsilon^{\alpha\dot\beta}\cald_{\alpha\dot\beta}\Phi \eqn\Jacob $$
where $\Phi$ stands for any matter field.

The action that results from \Daniel\ after putting $\lambda$, $\eta$, $\chi$
and $G$ to zero is invariant under the truncated transformations written above,
provided that the parameters $\epsilon$ and $\epsilon_{\alpha\dot\beta}$ are
covariantly constant. This action takes the following form,  $$  {\cal L} =
{\cal L}^{\hbox{\sevenrm DW}} + {\cal L}_0 + m{\cal L}_m, \eqn\Levy  $$ where,
$$ \eqalign{ {\cal L}_0 = & \int d^4x e  \Big[
\overline{H}^{\alpha}(\tableau{1}+{1\over 4}R) H_{\alpha} +{i\over
2}\overline{H}^\alpha F_{\alpha\beta}^{+} H^\beta +{i}
\overline{u}^{\alpha}\cald_{{\alpha\dot\alpha}}{v}^{\dot\alpha} -
{i}\overline{v}^{\dot\alpha}\cald_{{\alpha\dot\alpha}}u^{\alpha} \cr
&\,\,\,\,\,\,\,\,\,\,\,\,\,\,\,\,\,\,\,\,\,\,\,\,\,\,\,\,\,\,
+\overline{K'}^{\alpha}K_{\alpha}'
+\overline{H}^{\alpha}\psi_{\alpha\dot\beta}v^{\dot\beta}
-\overline{v}^{\dot\beta}\psi_{\alpha\dot\beta} H^{\alpha}+i\overline
v^{\dot\alpha}\phi v_{\dot\alpha}\Big],\cr
 m {\cal L}_m = &
 \int d^4x e \Big[m^2 \overline{H}^\alpha H_\alpha +m(\overline{u}^\alpha
u_\alpha + \overline{v}^{\dot\alpha}v_{\dot\alpha}) -i
m\overline{H}^{\alpha}\phi H_{\alpha} \Big].\cr} \eqn\Joshua $$

The action \Levy\ represents the coupling of topological matter to a subset of
the Donaldson-Witten multiplet. The question arises if this truncation can be
consistently extended to \diecisiete, \dieciocho, \diecinueve\ and \cauchy\ and
formulate a theory of topological matter coupled to topological Yang-Mills
which would be simpler than the preceding one. If we put $\lambda$, $\eta$,
$\chi$ and $G$ to zero in \diecisiete, \dieciocho, \diecinueve\ and \cauchy,
the resulting action contains only $F_{+}^{2}$, and is not $Q$-invariant. Then,
this procedure has to be discarded. Another possibility is simply to leave all
what concerns to Donaldson-Witten theory untouched, both the action and the
$Q$-transformations of its fields. Then there is no problem, and we arrive to a
satisfactory theory. But an important caveat should be pointed out. The
truncated
$Q$-transformations of the Donaldson-Witten fields coincide with the old ones,
and this is why the truncation can be extended, but the $G$ transformations are
different before and after the truncation, and this destroys the $G$ symmetry
of \cauchy.

The conclusion is that the truncation yields a $Q$-invariant theory of
topological matter coupled to topological Yang-Mills, but $G$-symmetry is lost.
If topological gravity is not at issue, this loose is irrelevant. Nevertheless,
the inclusion of topological gravity will probably need that symmetry, and then
singles out the whole theory as the only one to which it can be consistently
coupled.  As for the topological character of the theory, this feature depends
on the $Q$-exactness of its energy-momentum tensor. It should be fully
$Q$-exact off-shell because the action is not exact and therefore no equations
of motion can be used. These aspects of the theory are discussed
 in the next section.

\endpage
\chapter{Energy-momentum tensors}

An important issue to address in every topological field theory of Witten type
is the calculation of the energy-momentum tensor. These theories involve fields
of integer and semi-integer spin (bosons, fermions and ghosts) living in a
curved manifold  $\mani$, so we need to introduce a vierbein $e^{a\mu}$ and a
spin connection $\omega^{ab}_{\mu}$ to define semi-integer spin fields and
their spacetime covariant  derivatives. We assume that $\mani$ admits these
structures.

{}From now on we will indicate vector indices on which the twisted local
Lorentz
transformations \tres\ act ($flat$ or $tangent$ indices) by using letters from
the beginning of the Latin alphabet (a, b, \dots), and vector indices on which
local translations (general coordinate transformations) act ($curved$ or
$world$ indices) by letters from the middle of the Greek alphabet ($\mu$,
$\nu$, \dots). The vierbein converts one kind of indices into the other.

It is necessary to declare which position of the indices is considered as
fundamental, and which is the result of applying the metric tensor. This is
important in order to keep track of dependences on the metric. Our conventions
are: $$ \eqalign{ e^{a\mu},\quad \cald_{\mu} \longrightarrow \quad
&\rm{fundamental} \cr e^{a}_{\mu}=g_{\mu\nu}e^{a\nu}, \quad
\cald^{\mu}=g^{\mu\nu}\cald_{\nu}\longrightarrow \quad & \rm{derived} \cr.}
\eqn\Nabucodonosor $$ Pauli matrices are always defined in a locally inertial
system of reference, and according to our conventions they carry a tangent
index. The action of the spacetime covariant derivative on spin-1/2 fields
takes the form $$ \eqalign{
\cald_{\mu}H_{\alpha}=&\partial_{\mu}H_{\alpha}+\half\omega^{ab}_{\mu}
\big(\sigma_{ab}\big)_{\alpha}^{\,\,\,\,\beta}H_{\beta}-iA_\mu H_\alpha , \cr
\cald_{\mu}\overline{H}^{\alpha}=&\partial_{\mu}\overline{H}^{\alpha}
-\half\omega^{ab}_{\mu}
\overline{H}^{\beta} \big(\sigma_{ab}\big)_{\beta}^{\,\,\alpha}
+i \overline{H}^\alpha A_\mu,\cr
\cald_{\mu}v^{\dot\alpha}=&\partial_{\mu}v^{\dot\alpha}+\half\omega^{ab}_{\mu}
\big(\tilde\sigma_{ab}\big)^{\dot\alpha}_{\,\,\,\,\dot\beta}v^{\dot\beta}
-i A_\mu v^{\dot\alpha}, \cr
\cald_{\mu}\overline{v}_{\dot\alpha}=&\partial_{\mu}\overline{v}_{\dot\alpha}
-\half\omega^{ab}_{\mu}
\overline{v}_{\dot\beta}
\big(\tilde\sigma_{ab}\big)^{\dot\beta}_{\,\,\,\,\dot\alpha}
+i\overline{v}_{\dot\alpha}A_\mu.
\cr}
 \eqn\Asurbanipal
$$
where $\sigma_{ab}$ are the spin matrices defined in
the Appendix. The last identity needed is the variation of the spin connection
under a change in the vierbein: $$
\delta\omega^{ab}_{\mu}=\half\big[e^{[a|\rho}\delta e^{b]}_{[\rho ;\mu]} +
e^{[a|\rho}e^{b]\sigma}e_{c\mu}\delta e^{c}_{\rho ;\sigma}\big]. \eqn\Jerjes
$$ The energy-momentum tensor is defined as: $$ T^{\mu\nu}={e_{a}^{\mu}\over
\sqrt{g}}{\delta S\over\delta e_{a\nu}}, \eqn\Darius
$$
where we consider arbitrary
variations of the vierbein, not only those which lead to variations of the
metric tensor. The $T_{\mu\nu}$ corresponding to  ${\cal L}_0^\laga$ in
\borel\ reads,
 $$
\eqalign{ T^{\mu\nu}&={1\over
4}\Big[\cald_{\alpha\dot\alpha}\overline{H}
^{\alpha}\big(\tilde\sigma^{(\nu}\big)^{\dot\alpha\beta}
{\buildrel\leftrightarrow\over{\cald^{\mu)}}}H_{\beta}-\overline{H}^{\alpha}
\big(\sigma^{(\nu}\big)_{\alpha\dot\alpha}{\buildrel\leftrightarrow
\over{\cald^{\mu)}}}\cald^{\dot\alpha\beta}H_{\beta}\cr
&-i\overline{u}^{\alpha}\big(\sigma^{(\nu}\big)_{\alpha\dot\alpha}
{\buildrel\leftrightarrow\over{\cald^{\mu)}}}v^{\dot\alpha}+i\overline{v}
_{\dot\alpha}\big(\tilde\sigma^{(\nu)}\big)^{\dot\alpha\alpha}
{\buildrel\leftrightarrow\over{\cald^{\mu)}}}u_{\alpha}\Big]\cr & +\half
g^{\mu\nu}\Big[\overline{H}^{\alpha}\cald_{\alpha\dot\alpha}
\cald^{\dot\alpha\beta}H_{\beta}+\big(\cald^{\dot\alpha\beta}
\cald_{\alpha\dot\alpha}\overline{H}^{\alpha}\big)H_{\beta}
+i\overline{u}^{\alpha}{\buildrel\leftrightarrow\over\cald}_{\alpha\dot\alpha}
v^{\dot\alpha}-i\overline{v}_{\dot\alpha}
{\buildrel\leftrightarrow\over\cald}^{\dot\alpha\alpha}u_{\alpha}\cr &
-2\overline{K}^{\alpha}K_{\alpha}
-2\overline{H}^{\alpha}\psi_{\alpha\dot\alpha}v^{\dot\alpha}
+2\overline{v}^{\dot\alpha}\psi_{\dot\alpha\alpha}H^{\alpha}
+2i\overline{H}^{\alpha}\eta u_{\alpha} -i\overline{u}^{\alpha}\eta H_{\alpha}
-2i\overline{H}^{\alpha}\chi_{\alpha\beta}u^{\beta}\cr &
+2i\overline{u}^{\alpha}\chi_{\alpha\beta}H^{\beta}
+2i\overline{u}^{\alpha}\lambda u_{\alpha} +2i\overline{v}^{\dot\alpha}\phi
v_{\dot\alpha} -i\overline{H}^{\alpha}G_{\alpha\beta}H^{\beta}
+\overline{H}^{\alpha}\big{\{}\phi, \lambda \big{\}}H_{\alpha} \Big] \cr &
+\half\Big[\overline{H}^{\alpha}
\big(\sigma^{\mu\nu}\big)_{\alpha}^{\,\,\,\,\beta}\cald_{\beta\dot\alpha}
\cald^{\dot\alpha\gamma}H_{\gamma}-\cald^{\dot\alpha\beta}
\cald_{\alpha\dot\alpha}\overline{H}^{\alpha}
\big(\sigma^{\mu\nu}\big)_{\beta}^{\,\,\,\,\gamma}H_{\gamma}
-i\overline{v}_{\dot\alpha}
\big(\tilde\sigma^{\mu\nu}\big)^{\,\,\,\,\dot\alpha}_{\dot\beta}
\cald^{\dot\beta\beta}u_{\beta}\cr
&-i\cald^{\dot\alpha\alpha}\overline{v}_{\dot\alpha}
\big(\sigma^{\mu\nu}\big)_{\alpha}^{\,\,\,\,\beta}u_{\beta}
+i\overline{u}^{\alpha}
\big(\sigma^{\mu\nu}\big)_{\alpha}^{\,\,\,\,\beta}
\cald_{\beta\dot\gamma}v^{\dot\gamma}+
\cald_{\alpha\dot\alpha}\overline{u}^{\alpha}
\big(\tilde\sigma^{\mu\nu}\big)^{\,\,\,\,\dot\alpha}_{\dot\beta}
v^{\dot\beta}\Big]. \cr} \eqn\Artajerjes $$ This $T^{\mu\nu}$ is not
$Q$-exact, even in the free case. If we only consider matter, it can be
written as a $Q$-variation of some $\Lambda^{\mu\nu}$ plus something that
vanishes on shell. The general expression is
$$
\eqalign{
T^{\mu\nu}=&\big{\{} Q,
\Lambda^{\mu\nu}_{\hbox{\sevenrm{e,A}}}\big{\}}+g^{\mu\nu}
T_{\hbox{\sevenrm{e,A}}}\cr
&-\half\big[\overline{H}^{\alpha}\big(\sigma^{(\mu}\big)_{\alpha\dot\alpha}
\psi^{\nu)}v^{\dot\alpha}+\overline{v}_{\dot\alpha}
\big(\tilde\sigma^{(\mu}\big)^{\dot\alpha\alpha}
\psi^{\nu)}H_{\alpha}+\overline{H}^{\alpha}
\big(\sigma^{\mu\nu}\big)_{\alpha}^{\,\,\,\,\beta}
\psi_{\beta\dot\beta}v^{\dot\beta}\cr
&-\overline{v}_{\dot\alpha}\big(\tilde\sigma^{\mu\nu}\big)^{\dot\alpha}
_{\,\,\,\,\dot\beta}\psi^{\dot\beta\beta}H_{\beta}+
\overline{H}^{\alpha}\psi_{\alpha\dot\alpha}\big(\tilde
\sigma^{\mu\nu}\big)^{\dot\alpha}_{\,\,\,\,\dot\beta}
v^{\dot\beta}-\overline{v}_{\dot\alpha}
\psi^{\dot\alpha\alpha}\big(\sigma^{\mu\nu}\big)_{\alpha}^{\,\,\,\,\beta}
H_{\beta}\big].
\cr} \eqn\Idomeneo $$ where
$$ \eqalign{
\Lambda^{\mu\nu}_{\hbox{\sevenrm{e,A}}}&=-{i\over4}\Big[\overline{H}^{\alpha}
\big(\sigma^{(\nu}\big)_{\alpha\dot\alpha}
{\buildrel\leftrightarrow\over{\cald^{\mu)}}}v^{\dot\alpha}+
\overline{v}_{\dot\alpha}\big(\tilde\sigma^{(\nu}\big)^{\dot\alpha\alpha}
{\buildrel\leftrightarrow\over{\cald^{\mu)}}}H_{\alpha}\Big]
+{i\over2}\Big[\overline{H}^{\alpha}
\big(\sigma^{\mu\nu}\big)_{\alpha}^{\,\,\,\,\beta}
\cald_{\beta\dot\beta}v^{\dot\beta}\cr
&+\overline{v}_{\dot\alpha}\big(\tilde\sigma^{\mu\nu}\big)
^{\dot\alpha}_{\,\,\,\,\dot\beta}\cald^{\dot\beta\beta}
H_{\beta}+\cald_{\alpha\dot\alpha}\overline{H}^{\alpha}
\big(\tilde\sigma^{\mu\nu}\big)^{\dot\alpha}_{\,\,\,\,\dot\beta}
v^{\dot\beta}+\cald^{\dot\alpha\alpha}
\overline{v}_{\dot\alpha}\big(\sigma^{\mu\nu}\big)_{\alpha}^{\,\,\,\,\beta}
H_{\beta}\Big],
\cr T_{\hbox{\sevenrm{e,A}}}&=-\half
\Big[\overline{H}^{\alpha}\cald_{\alpha\dot\alpha}
\cald_{\beta}^{\,\,\dot\alpha}H^{\beta}+\big(\cald_{\beta}^{\,\,\dot\alpha}
\cald_{\alpha\dot\alpha}\overline{H}^{\alpha}\big)H^{\beta}
-i\overline{u}^{\alpha}{\buildrel\leftrightarrow\over\cald}_{\alpha\dot\alpha}
v^{\dot\alpha}+i\overline{v}^{\dot\alpha}
{\buildrel\leftrightarrow\over\cald}_{\dot\alpha\alpha}u^{\alpha}\cr &
-2\overline{K}^{\alpha}K_{\alpha}-2\overline{H}^{\alpha}
\psi_{\alpha\dot\alpha}v^{\dot\alpha}
+2\overline{v}^{\dot\alpha}\psi_{\dot\alpha\alpha}H^{\alpha}
+2i\overline{H}^{\alpha}\eta u_{\alpha} -i\overline{u}^{\alpha}\eta H_{\alpha}
-2i\overline{H}^{\alpha}\chi_{\alpha\beta}u^{\beta}\cr &
+2i\overline{u}^{\alpha}\chi_{\alpha\beta}H^{\beta}
+2i\overline{u}^{\alpha}\lambda u_{\alpha} +2i\overline{v}^{\dot\alpha}\phi
v_{\dot\alpha} -i\overline{H}^{\alpha}G_{\alpha\beta}H^{\beta}
+\overline{H}^{\alpha}\big{\{}\phi, \lambda \big{\}}H_{\alpha}\Big].\cr}
\eqn\Telemaco $$ In the free case the last two sets of terms in \Idomeneo\ do
not appear and therefore the energy-momentum tensor, excluding the part
proportional to $g_{\mu\nu}$, is $Q$-exact. On the other hand, the part
proportional to $g_{\mu\nu}$ vanishes on-shell. Since in the free case the
action is $Q$-exact we are allowed to take the energy-momentun tensor on-shell
and we therefore conclude that the theory is topological. The situation is
very different for the theory coupled to topological Yang-Mills. Either the
full or the truncated theory do not possess neither a $Q$-exact action nor a
$Q$-exact energy-momentum tensor. The energy momentum tensor is $Q$-exact up
two terms which vanish on-shell. However, since in this case the theory is not
necessarily exact in the small coupling limit we can not conclude that the
theory is topological. These facts are an indication that this theory might
represent  a new phenomena related to a  topological symmetry breaking caused
by the introduction of matter interactions.

The calculation of the energy-momentum tensor of \cauchy\ does not offer any
new difficulty. It turns out to be $Q$-exact, as a consequence of the
$Q$-exactness of the  action and the independence of the transformations
\diecisiete\ on the underlying geometry. Notice that the term
$F^{+}_{\alpha\beta}$ is a component of a two-form. The transformation of
$\psi_{\alpha\dot\alpha}$ include a  covariant derivative, but it acts on a
scalar. Finally, the only field that transforms with the covariant derivative
of a vector is $G_{\alpha\beta}$, which is an auxiliary field that can be set
to zero.

The conclusion is that neither the full nor the truncated theories are strictly
topological, since the action is $Q$-invariant, but not $Q$-exact, and the
energy-momentum tensor is $Q$-exact only on-shell. As the theory does not
coincide with its classical limit, the non $Q$-exact terms in $T_{\mu\nu}$
cannot be discarded.

\endpage
\chapter{Another type of topological matter in 4D}

So far all our models have been built starting from an $N=2$ theory and
performing a twist. One can also think of starting directly from a set of
fields which satisfy some $\delta$-transformation properties, and try to write
an action invariant under that transformations. The only drawback of  this
approach is the difficulty to implement other symmetries (as $\delta'$ or
$\delta''$). We present now an example of this procedure, which turns out to be
a truncated twisted version of the relaxed hypermultiplet [\howe]. The analysis
to build this model from the relaxed hypermultiplet will not be presented here.
It goes along the same lines as the ones which led to the matter models in the
previous sections. We present only the truncated model and therefore all
symmetries but the one corresponding to $Q$ have been lost. A coupling of this
model to topological gravity would need the full theory. Its form will
presented elsewhere.

The basic set of fields of the model has the same spin content as in the
previous model. We will denote these fields by
$H_\alpha, u_\alpha, v_{\dot\alpha}$, and $L_{\dot\alpha}$. The reasons to
make some of the choices done for the previous model will become clear below.
The symmetry transformations turn out to be:
$$
\eqalign{
\deltat H_{\alpha} &=u_{\alpha}, \cr \deltat u_{\alpha}&=0, \cr \deltat
v^{\dot\alpha}
&=L^{\dot\alpha}-\D_{\alpha}^{\,\,\,\,\,\dot\alpha}H^{\alpha}      , \cr
\deltat L^{\dot\alpha}&=\D_{\alpha}^{\,\,\,\,\,\dot\alpha}u^{\alpha}
,\cr} \qquad \eqalign{ \deltat \overline{H}_{\alpha} &=\overline{u}_{\alpha},
\cr \deltat \overline{u}_{\alpha}&=0, \cr \deltat
\overline{v}^{\dot\alpha}&=\overline{L}^{\dot\alpha}
-\D_{\alpha}^{\,\,\,\,\,\dot\alpha}\overline{H}^{\alpha}    , \cr  \deltat
\overline{L}^{\dot\alpha}
&=\D_{\alpha}^{\,\,\,\,\,\dot\alpha}\overline{u}^{\alpha}. \cr} \eqn\KingDavid
$$ Note that this $\deltat$ is nilpotent, which considerably simplifies the
issue of writing invariant actions; it suffices to look for $\deltat$-exact
functionals of the fields with the right dimension and ghost number. The field
$L_{\dot\alpha}$ has the right dimension to be an auxiliary field, and
it can be thought
as the field needed to render the $\deltat$-transformation nilpotent. The
next task is to find out  the action which is annihilated by this
transformation. The choice which leads to a $\deltat$-exact action with
kinetic terms and  includes
$L_{\dot\alpha}$ as an auxiliary field is the following:
$$
{\cal L}_2 = \half \deltat
\int_{M} d^4x \sqrt{g}\Big[ \overline{H}^\alpha \D_{\alpha\dot\alpha}
v^{\dot\alpha} +\overline{v}^{\dot\alpha}\D_{\alpha\dot\alpha}H^{\alpha}
+\overline{L}^{\dot\alpha} v_{\dot\alpha}
+\overline{v}^{\dot\alpha}L_{\dot\alpha}\Big]. \eqn\Washington
$$
Its full
expanded expression reads almost exactly like the one of our previous model
\maria\ in the massless case:
$$ {\cal L}_2  = \int d^4x
\sqrt{g}\Big[\D_{\alpha\dot\beta}\overline{H}^{\alpha}
\D_{\beta}^{\,\,\,\,\,\dot\beta}
H^{\beta}+\overline{u}^{\alpha}\D_{\alpha\dot\alpha}v^{\dot\alpha}-
\overline{v}^{\dot\alpha}\D_{\dot\alpha\alpha}u^{\alpha}
+\overline{L}^{\dot\alpha}{L}_{\dot\alpha} \Big].   \eqn\Jefferson
$$ Being
the action ${\cal L}_2$ manifestly $\deltat$-invariant, the only
condition for the theory being topological is the $\deltat$-exactness of its
energy-momentum tensor. This is a highly non trivial condition since, as we
have seen in the previous chapter, the exactness of the action does not imply
the same property for the energy-momentum tensor because of the dependence of
the transformation on the spin connection through the covariant derivatives.
In our case some help to end with a topological quantum field theory comes from
the fact that we can effectively consider
 the theory on-shell due to the exactness of the
action, as we did in our previous model in the free case. We show now that
this is also the case in the present model. The energy-momentum tensor can be
written as follows: $$ T_{\mu\nu}=\deltat\Lambda_{\mu\nu}+M_{\mu\nu},
\eqn\Madison $$ where, $$
\eqalign{ \Lambda^{\mu\nu}&={1\over4}\Big[\overline{H}^{\alpha}
\big(\sigma^{(\nu}\big)_{\alpha\dot\alpha}
{\buildrel\leftrightarrow\over{\D^{\mu)}}}v^{\dot\alpha}+
\overline{v}_{\dot\alpha}\big(\tilde\sigma^{(\nu}\big)^{\dot\alpha\alpha}
{\buildrel\leftrightarrow\over{\D^{\mu)}}}H_{\alpha}\Big]
-\half\Big[\overline{H}^{\alpha}
\big(\sigma^{\mu\nu}\big)_{\alpha}^{\,\,\,\,\beta}
\D_{\beta\dot\beta}v^{\dot\beta}\cr
&+\overline{v}_{\dot\alpha}\big(\tilde\sigma^{\mu\nu}\big)
^{\dot\alpha}_{\,\,\,\,\dot\beta}\D^{\dot\beta\beta}H_{\beta}+
\D_{\alpha\dot\alpha}\overline{H}^{\alpha}
\big(\tilde\sigma^{\mu\nu}\big)^{\dot\alpha}_{\,\,\,\,\dot\beta}v^{\dot\beta}
+\D^{\dot\alpha\alpha}
\overline{v}_{\dot\alpha}\big(\sigma^{\mu\nu}\big)_{\alpha}^{\,\,\,\,\beta}
H_{\beta}\Big]
\cr &+\half
g^{\mu\nu}\Big[\overline{L}_{\dot\alpha}v^{\dot\alpha}+
\overline{v}_{\dot\alpha}L^{\dot\alpha}
+\overline{v}^{\dot\alpha}\D_{\alpha\dot\alpha}H^{\alpha}-
\overline{H}^{\alpha}\D_{\alpha\dot\alpha}
v^{\dot\alpha}\Big],\cr
M_{\mu\nu}&={1\over4}\Big[\overline{H}^{\alpha}
\big(\sigma_{(\mu}\big)_{\alpha\dot\beta}
\D_{\nu)}L^{\dot\beta}+\D_{(\mu}L_{\dot\alpha}
\big(\tilde\sigma_{\nu)}\big)^{\dot\alpha}_{\,\,\,\,\alpha}H^{\alpha}\cr
&-\D_{(\mu}
\overline{H}^{\alpha}\big(\sigma_{\nu)}\big)_{\alpha\dot\beta}
L^{\dot\beta}-L_{\dot\alpha}
\big(\tilde\sigma_{(\mu}\big)^{\dot\alpha}_{\,\,\,\,\alpha}
\D_{\nu)}H^{\alpha}\Big]
\cr&+\half\Big[\overline{L}_{\dot\alpha}
\big(\tilde\sigma_{\mu\nu}\big)^{\dot\alpha}_{\,\,\,\,\dot\beta}
\D_{\alpha}^{\,\,\,\,\dot\beta}H^{\alpha}+
\D^{\gamma}_{\,\,\,\,\dot\alpha}\overline{L}^{\dot\alpha}
\big(\sigma_{(\mu\nu}\big)_{\gamma}^{\,\,\,\,\alpha}H_{\alpha}-
\D_{\alpha\dot\beta}\overline{H}^{\alpha}
\big(\tilde\sigma_{\mu\nu}\big)^{\dot\beta}_{\,\,\,\,\dot\gamma}L^{\dot\gamma}
\cr &-\overline{H}^{\alpha}
\big(\sigma_{\mu\nu}\big)_{\alpha}^{\,\,\,\,\beta}
\D_{\beta\dot\gamma}L^{\dot\gamma}\Big]
-\half g_{\mu\nu}\Big[\overline{H}^{\alpha}\D_{\alpha\dot\beta}L^{\dot\beta}+
\big(\D_{\alpha}^{\,\,\,\,\dot\alpha}L_{\dot\alpha}\big)H^{\alpha}\Big].
 \cr} \eqn\Lincoln $$ The crucial observation is that, as explained before, we
can set the auxiliary field to zero. In this case the energy-momentum tensor is
$\deltat$-exact and therefore this theory is topological. Now we define a
coupling of this multiplet to the Donaldson-Witten multiplet by means of the
following generalized transformations:
 $$ \eqalign{ \deltat H_{\alpha}
&=u_{\alpha}, \cr \deltat u_{\alpha}&=-i\phi H_{\alpha}, \cr \deltat
v^{\dot\alpha}
&=L^{\dot\alpha}-\cald_{\alpha}^{\,\,\,\,\,\dot\alpha}H^{\alpha}      , \cr
\deltat L^{\dot\alpha}&=\cald_{\alpha}^{\,\,\,\,\,\dot\alpha}u^{\alpha}
-i\psi_{\alpha}^{\,\,\,\,\,\dot\alpha} H^{\alpha} -i\phi v^{\dot\alpha}
,\cr} \qquad \eqalign{ \deltat \overline{H}_{\alpha} &=\overline{u}_{\alpha},
\cr \deltat \overline{u}_{\alpha}&=i\phi \overline{H}_{\alpha}, \cr \deltat
\overline{v}^{\dot\alpha}&=\overline{L}^{\dot\alpha}
-\cald_{\alpha}^{\,\,\,\,\,\dot\alpha}\overline{H}^{\alpha}    , \cr  \deltat
\overline{L}^{\dot\alpha}
&=\cald_{\alpha}^{\,\,\,\,\,\dot\alpha}\overline{u}^{\alpha}
+i\psi_{\alpha}^{\,\,\,\,\,\dot\alpha} \overline{H}^{\alpha} +i\phi
\overline{v}^{\dot\alpha}  .\cr} \eqn\Roosevelt
$$
 As in Donaldson-Witten
theory, these transformations close up to a gauge transformation whose gauge
parameter is $\phi$: $$ \eqalign{ \deltat^2 H_{\alpha}&= -i\phi H_{\alpha}, \cr
\deltat^2  u_{\alpha}&= -i\phi u_{\alpha}, \cr \deltat^2  v^{\dot\alpha} &=
-i\phi v^{\dot\alpha}, \cr \deltat^2  L^{\dot\alpha} &= -i\phi L^{\dot\alpha}
.\cr} \eqn\Kennedy $$
This implies that  we can take any suitable quantity of ghost
number -1 and obtain an  action using $\deltat$. Such a quantity just must be
gauge invariant. The free one does the job, and we would obtain the same
action and energy-momentum tensor as in the truncated theory except  for the
$K^\alpha K_\alpha$ term, which is now $-L^{\dot\alpha}L_{\dot\alpha}$:
$$ \tilde S^\lag = {\cal L}_\ym^\lag + \tilde
{\cal L}^\laga_0, \eqn\borelino $$ where,
$$
\eqalign{ \tilde {\cal L}^\laga_0
= & \int d^4x e  \Big[ \overline{H}^{\alpha}(\tableau{1}+{1\over 4}R)
H_{\alpha} +{i\over 2}\overline{H}^\alpha F_{\alpha\beta}^{+} H^\beta +{i}
\overline{u}^{\alpha}\cald_{{\alpha\dot\alpha}}{v}^{\dot\alpha} -
{i}\overline{v}^{\dot\alpha}\cald_{{\alpha\dot\alpha}}u^{\alpha} \cr
&\,\,\,\,\,\,\,\,\,\,\,\,\,\,\,\,\,\,\,\,\,\,\,\,\,\,\,\,\,\,
-\overline{L}^{\alpha}L_{\alpha}'
+\overline{H}^{\alpha}\psi_{\alpha\dot\beta}v^{\dot\beta}
-\overline{v}^{\dot\beta}\psi_{\alpha\dot\beta} H^{\alpha}+i\overline
v^{\dot\alpha}\phi v_{\dot\alpha}\Big].\cr}
\eqn\Joshuados
$$ Actually, one
has a result entirely analogue to the one obtained in  the free case,
\Madison\ and \Lincoln. The energy-momentum tensor is then $\deltat$-exact in
the coupled case except for terms linear in the auxiliary field. This fact,
together with the exactness of the action leads to the conclusion that this
theory is topological in both the free and the coupled case.

\endpage
\chapter{Final comments and remarks}

Let us first analyze the features of the possible observables of the models
presented in the previous sections.  Observables are $Q$-invariant quanties
constructed out of the fields of the theory. Certainly, the observables
\observables\ of topological Yang-Mills theory are $Q$-invariant quantities
since all the fields entering in them possess the same $Q$-transformations
before and after the coupling to matter fields. One would like to have another
set of observables  involving matter fields. We have done a thorough analysis
to find observables which involve matter fields and we have not found any. This
analysis goes in two steps. First, one writes all possible gauge invariant
operators quadratic in matter fields of a given ghost number. Then one checks
if it is possible to obtain a linear combination of them which is $Q$-invariant
and it is not $Q$-exact. Our analysis shows that there are no operators a that
type. Considering powers of these operators one is led the same conclusion.
 One is therefore left to the study of the observables \observables\ in the
presence of matter. Of course, the resulting vev of the operators \observables\
are rather different than in the theory with no matter. There are relevant
contributions from the matter fields in their functional integration.  These
contributions are very important. Indeed, for the models of sect. 6 and 7 it
is not guaranteed that these observables lead to topological quantitites. This
follows from the fact that both, the action and  energy-momentum tensor of the
theory, are not $Q$-exact. For the topological matter model of sect. 8,
however, since the action is $Q$-exact, the small coupling limit is exact and
the observables do indeed lead to topological quantities.

 The computation of these observables using the action \borel\ leads to
quantities which are labeled, besides the usual homology cycles,  by the matter
representation chosen in the action \borel. These quantities have properties
very similar to Donaldson invariants. Let us denote by $H_*(M)$ the homology
groups of the spin manifold $M$. These observables are also polynomials on
$H_*(M)\times H_*(M) \times \dots \times H_*(M)$ as Donaldson invariants are.
This property is based only on the fact that the exterior differential of any
of the operators \obs\ is $Q$-exact. There is no need to have a $Q$-exact
energy-momentum tensor for this to hold, just a $Q$-invariant action as it is
the case. On the other hand, these observables, as in Donaldson-Witten theory,
can be computed in the limit $e\rightarrow 0$. The reason for this is that all
the dependence on the coupling constant $e$ in \borel\ and \borelino\ is
contained in a part of the action which is $Q$-exact. Again, there is no need
of a $Q$-exact energy-momentum tensor for this to hold. This implies in
particular that the vev of the observables are independent of $e$. Thus, the
quantities associated to the vev of arbitrary products of the operators
\observables\ constitute a generalization of Donaldson invariants which,
however, it is not guaranteed that in general are topological invariants.
The breaking of the invariance comes from the fact for an arbitrary product of
operators \observables,  $$ {2\over\sqrt{g}}{\delta\over \delta g^{\mu\nu} }
\langle \prod {\cal O}^{(\gamma)} \rangle = \langle \prod {\cal O}^{(\gamma)}
\tau_{\mu\nu} \rangle, \eqn\lae $$  where $\tau_{\mu\nu}$ is the part of the
energy-momentum tensor which is not $Q$-exact. For models based on the action
\borel\ one does not possess an argument ensuring that \lae\ vanishes. For
models based on the action \borelino, however,  one can argue that the right
hand side  of \lae\ vanishes using the $Q$-exactness of the action.

It is also important to remark that the actions \borel\ and \borelino\ of the
models under considerations have a very similar structure. Their difference is
very subtle since it resides in the form of the auxiliary fields. One would
have to study in detail the role played by the auxiliary fields. Two
possibilities could occur. If their role is trivial, the model based on the
action
\borel\ would be equivalent to the model based on the action \borelino\ and
therefore one could conclude that the model based on \borel\
is topological. If the role
played by the auxiliary fields is non-trivial the model based on the action
\borel\ could very well represent a situation in which the topological
symmetry is broken.
In either case that model is interesting and deserves
further investigations.

Let us consider finally the question of the mirror-like behavior in four
dimensions. The two models which have been constructed do not seem to lead to
this kind of phenomena. Their difference resides on the auxiliary fields and
these possess rather simple couplings. It is likely that one has to study the
coupling of the two types of matter to topological gravity to observe some
kind of mirror-like phenomena.  Certainly, the structure of the couplings of
the auxiliary fields of the matter multiplets will be much more
complicated. This study, however, requires to build the full theory resulting
from the twisting of the relaxed hypermultiplet, and not the truncated one
presented in sect. 8. An additional set of auxiliary fields appear in that
situation which announces that the analysis of the models are rather
different. Matter couplings to topological gravity will be studied in future
works.

\vskip1cm

\ack We would like to thank A. V. Ramallo for very helpful discussions.  This
work was supported in part by DGICYT under grant PB90-0772 and by CICYT under
grants AEN93-0729 and AEN94-0928.

\endpage


\appendix

In this work we use spinor notation for all Lorentz representations, denoting
spinor indices by Greek letters, dotted for $(0,\half)$, and undotted for
$(\half,0)$. An arbitrary irreducible representation $(L,R)$ is labeled with
$2L$ totally symmetrized undotted indices, and $2R$ totally symmetrized dotted
indices. The symmetrization symbol of $N$ indices means the sum over all their
permutations without $1/N!$, and similarly for antisymmetrization. For
instance: $$\eqalign{ X_{(\alpha\beta)}=&X_{\alpha\beta}+X_{\beta\alpha},\cr
X_{[\alpha\beta]}=&X_{\alpha\beta}-X_{\beta\alpha},\cr} \eqn\symmetry $$ and
similarly for dotted indices. Spinor indices are raised and lowered by the
second-rank antisymmetric symbol $C_{\alpha\beta}$: $$
\psi_{\alpha}=\psi^{\beta}C_{\beta\alpha},\qquad
\psi^{\alpha}=C^{\alpha\beta}\psi_{\beta}.
\eqn\indices $$ The same
convention holds for dotted indices. The $C_{\alpha\beta}$ symbol is defined
as follows: $$ C_{\alpha\beta}=\pmatrix{0&-i\cr
i&0\cr}=C^{\dot\alpha\dot\beta}, \qquad C^{\alpha\beta}=\pmatrix{0&i\cr
-i&0\cr}=C_{\dot\alpha\dot\beta}.
 \eqn\matrices
$$ Vectors belong to the
$(\half,\half)$ representation, and are labeled with one undotted and one
dotted index. We have to distinguish two kinds of vectors: commuting and
anticommuting. We underline the composite index in the case of commuting
vectors. Examples of these two types of vectors are
$A_{\underline{\alpha\dot\beta}}$ and $\psi_{\alpha\dot\beta}$. Tangent vector
indices are denoted with lower-case roman letters, and are related to the
$\alpha\dot\beta$ basis by some Clebsch-Gordan coefficients, the Pauli
matrices,
 $$
 X^{\alpha\dot\beta}=\sigma_{a}^{\,\,\,\alpha\dot\beta} X^a \qquad
X^a =\half\sigma_{\alpha\dot\beta}^{\,\,\, a} X^{\alpha\dot\beta}.
\eqn\Diomedes
$$
Pauli matrices with world vector indices are defined by means
of the vierbein:
$$
\big(\sigma_{\mu}\big)_{\alpha\dot\alpha}=
e^a_{\mu}\big(\sigma_a\big)_{\alpha\dot\alpha},
\qquad
\big(\tilde\sigma_{\mu}\big)^{\dot\alpha\alpha}=e^a_{\mu}
\big(\tilde\sigma_a\big)^{\dot\alpha\alpha}.
\eqn\Ayax
$$
The vierbein satisfies the usual relations
$$
e_a^{\mu}e^{a\nu}=g^{\mu\nu},\qquad e_a^{\mu}e_{b\mu}=\eta_{ab},
 \eqn\Diomedesprima
$$ where $\eta_{ab}$ is the flat minkowskian metric tensor. The vierbein is
required to be covariantly constant: $$ \D_{\mu}e^{a\nu}=0. \eqn\Ulises $$ This
last identity relates the vierbein to the spin connection: $$
\omega^{ab}_{\mu}=\half\big[e^{[a|\rho} e^{b]}_{[\rho ;\mu]} +
e^{[a|\rho}e^{b]\sigma}e_{c\mu} e^{c}_{\rho ;\sigma}\big]. \eqn\Menelao $$
Although the spin connection does not transform as a tensor, its variation
does. This property permits to calculate $\delta\omega^{ab}_{\mu}$ simply going
to a locally inertial system of reference and covariantizing the result.
Its variation is given in \Darius.

Pauli matrices satisfy the following identities  $$ \eqalign{
\big(\tilde\sigma^{\mu}\sigma^{\nu} \big)^{\dot\alpha}_{\,\,\,\,\dot\beta}=&
g^{\mu\nu}\delta_{\dot\beta}^{\,\,\,\,\dot\alpha}+2\big(\tilde\sigma^{\mu\nu}
\big)^{\dot\alpha}
_{\,\,\,\,\dot\beta},\cr \big(\sigma^{\mu}\tilde\sigma^{\nu}
\big)_{\alpha}^{\,\,\,\,\beta}=&
g^{\mu\nu}\delta_{\alpha}^{\,\,\,\,\beta}+2\big(\sigma^{\mu\nu}\big)_{\alpha}
^{\,\,\,\,\beta}.\cr} \eqn\Paris $$ which also serve to define the spin
matrices $\sigma^{\mu\nu}$ and $\tilde\sigma^{\mu\nu}$. These matrices are
antisymmetric in their vector indices and symmetric in their spin indices.
Some useful identities  needed in the calculation of the energy-momentum
tensor are:
$$
\eqalign{
\big(\sigma^{\mu\nu}\big)_{\alpha}^{\,\,\,\,\beta}
\big(\sigma^{\lambda}\big)_{\beta\dot\beta}=&-\half
g^{\lambda[\mu}\big(\sigma^{\nu]}\big)_{\alpha\dot\beta}+{i\over2}
\epsilon^{\mu\nu\lambda\kappa}
\big(\sigma_{\kappa}\big)_{\alpha\dot\beta},\cr
\big(\tilde\sigma^{\mu\nu}\big)^{\dot\alpha}_{\,\,\,\,\dot\beta}
\big(\tilde\sigma^{\lambda}\big)^
{\dot\beta\beta}=&-\half
g^{\lambda[\mu}\big(\tilde\sigma^{\nu]}\big)^{\dot\alpha\beta}-{i\over2}
\epsilon^{\mu\nu\lambda\kappa}
\big(\tilde\sigma_{\kappa}\big)^{\dot\alpha\beta},\cr
\big(\sigma^{\lambda}\big)_{\alpha\dot\alpha}\big(\tilde\sigma^{\mu\nu}
\big)^{\dot\alpha}
_{\,\,\,\,\dot\beta}=&\half
g^{\lambda[\mu}\big(\sigma^{\nu]}\big)_{\alpha\dot\beta}+
{i\over2}\epsilon^{\mu\nu\lambda\kappa}
\big(\sigma_{\kappa}\big)_{\alpha\dot\beta},\cr
\big(\tilde\sigma^{\lambda}\big)^{\alpha\dot\alpha}
\big(\sigma^{\mu\nu}\big)_{\alpha}
^{\,\,\,\,\beta}=&\half
g^{\lambda[\mu}\big(\tilde\sigma^{\nu]}\big)^{\dot\alpha\beta}-{i\over2}
\epsilon^{\mu\nu\lambda\kappa}
\big(\tilde\sigma_{\kappa}\big)^{\dot\alpha\beta}.\cr} \eqn\Aquiles $$
Self-dual tensors in four dimensions have two symmetric undotted indices. The
self-dual part of an antisymmetric second-order tensor can be extracted using
the spin matrices, $\sigma^{ab}$:  $$ X_{\alpha\beta}^{+}=
2X^{ab}\big(\sigma_{ab}\big)_{\alpha\beta}, \qquad
X_{\dot\alpha\dot\beta}^{-}=2X^{ab}\big(\tilde\sigma_{ab}
\big)_{\dot\alpha\dot\beta}.
\eqn\Hector
$$
\endpage
\refout
\end